\documentclass{aa}  

\usepackage{graphicx}
\usepackage[colorlinks=true,
    linkcolor=blue,
    citecolor=blue,
    filecolor=magenta,      
    urlcolor=cyan]{hyperref}
\usepackage{booktabs}
\usepackage{natbib}
\usepackage{caption}
\usepackage{subcaption}
\usepackage{comment}
\usepackage{appendix}
\graphicspath{ {figures/} }
\usepackage{txfonts}
\usepackage{float}
\newcommand*{\unit}[1]{\ensuremath{\mathrm{\,#1}}}
\usepackage[normalem]{ulem}
\usepackage[dvipsnames]{xcolor}
\usepackage{afterpage}
\usepackage{bbding}
\usepackage{pifont}
\usepackage{afterpage}
\usepackage{bm}
\newcommand{\angstrom}{\mbox{\normalfont\AA}}
\usepackage{orcidlink}

\newcommand{\soutPC}{\bgroup\markoverwith{\textcolor{cyan}{\rule[0.5ex]{2pt}{1pt}}}\ULon}

\begin{document}

\title{Enhanced activity in close dual-AGN systems in the local Universe}

\author{Lorenzo~Battistini\inst{1,2}\fnmsep\thanks{Corresponding author, \email{lorenzo.battistini@inaf.it}}\orcidlink{0009-0000-4901-7987} \and Alessandra~De~Rosa\inst{2}\orcidlink{0000-0001-5668-6863} \and Paola~Severgnini\inst{3}\orcidlink{0000-0001-5619-5896} \and Cristian~Vignali\inst{4,5} \and
Jasbir~Singh\inst{3} \and
Pedro~R.~Capelo\inst{6}\orcidlink{0000-0002-1786-963X} \and 
Elena~Bertola\inst{7}\orcidlink{0000-0001-5487-2830} \and Stefano~Bianchi\inst{1} \and Quirino~D'Amato\inst{7} \and Matteo~Guainazzi\inst{8} \and 
Fabio~La~Franca\inst{1} \and 
Isabella~Lamperti\inst{9,7}\orcidlink{0000-0003-3336-5498}  \and 
Filippo~Mannucci\inst{7} \and Manali~Parvatikar\inst{2} \and Enrico~Piconcelli\inst{10} \and Federica~Ricci\inst{1} \and 
Fabio~Rigamonti\inst{3,11,12} \and 
Martina~Scialpi\inst{13,9,8}\orcidlink{0009-0006-5100-4986} \and Maria Vittoria~Zanchettin\inst{7}}
\authorrunning{L. Battistini et al}
\titlerunning{Dual-AGN triggering in the local Universe}

\institute{{Dipartimento di Matematica e Fisica, Università degli Studi Roma Tre, via della Vasca Navale 84, 00146 Roma, Italy}
\and {INAF - Istituto di Astrofisica e Planetologia Spaziali (IAPS), via Fosso del Cavaliere, Roma, 1-133, Italy}
\and {INAF - Osservatorio Astronomico di Brera, via Brera 28, Milano, 20141, Italy}
\and {Dipartimento di Fisica e Astronomia ‘Augusto Righi’, Università degli Studi di Bologna, via Gobetti 93/2, I-40129 Bologna, Italy}
\and {INAF – Osservatorio di Astrofisica e Scienza dello Spazio di Bologna, Via Gobetti 93/3, I-40129 Bologna, Italy}
\and {Department of Astrophysics, University of Zurich, Winterthurerstrasse 190, CH-8057 Z{\"u}rich, Switzerland}
\and {INAF - Osservatorio Astrofisico di Arcetri, Largo E. Fermi 5, I-50125 Firenze, Italy}
\and {European Space Agency (ESA), European Space Research and Technology Centre (ESTEC), Keplerlaan 1, 2201 AZ Noordwijk, The Netherlands}
\and {Università di Firenze, Dipartimento di Fisica e Astronomia, via G. Sansone 1, I-50019 Sesto F.no, Firenze, Italy}
\and {INAF Osservatorio Astronomico di Roma, Via Frascati 33, 00078 Monte Porzio Catone (RM), Italy}
\and {INFN, Sezione di Milano-Bicocca, Piazza della Scienza 3, I-20126 Milano, Italy}
\and {Como Lake centre for AstroPhysics (CLAP), DiSAT, Università dell’Insubria, via Valleggio 11, 22100 Como, Italy}
\and {University of Trento, Via Sommarive 14, I-38123 Trento, Italy}
}

\date{}

\abstract
{We present the study of an X-ray selected sample of active galactic nuclei (AGN) in pairs at projected spatial separations $1<r_{\rm p}/\unit{kpc}<100$ at $z<0.1$, using XMM-Newton and Chandra data. The pair sample is derived from an initial pool of approximately $2,000$ X-ray-selected AGN, and is composed of both AGN-AGN pairs (so called dual AGN) and AGN-galaxy pairs. From this selection, we find that approximately $10\%$ of AGN reside in pairs, and about $4\%$ are paired with another AGN. We performed a detailed X-ray and SDSS optical spectral analysis for AGN in duals and X-ray analysis for AGN in AGN-galaxy pairs, to characterise their absorption properties and investigate the possible triggering mechanisms. We then investigated how obscuration, luminosity, and Eddington ratio depend on projected separation $r_{\rm p}$. Amongst all  AGN in pairs, we found that $\sim$55\% are obscured (with hydrogen column density $N_\mathrm{H}>10^{22}\unit{cm^{-2}}$), amongst which $\sim$6\% are Compton-thick  ($N_{\rm H}>10^{24}\unit{cm^{-2}}$). The  fraction of absorbed AGN  is significantly higher in late-stage mergers ($r_{\rm p}<30\unit{kpc}$) compared to early-stage mergers ($r_\mathrm{p}>60\unit{kpc}$). Amongst the AGN in pairs,  we also observed an average excess of AGN pairs with respect to a control sample of inactive galaxies in pairs, and that such excess significantly increases with decreasing {$r_\mathrm{p}$} only for obscured AGN. Finally, in dual-AGN systems, both the bolometric luminosity and the Eddington ratio of the less massive black hole in the pair increase as the separation decreases. These findings suggest that mergers may have an important role in triggering AGN accretion and activity.}


   \maketitle
   
   \nolinenumbers

\section{Introduction}\label{sec.1}

The search and study of dual active galactic nuclei (dual AGN, with projected spatial separation between nuclei $r_{\rm p}\sim100\unit{pc}$--$100\unit{kpc}$) is a topic of great interest in modern astrophysics. Dual AGN are in fact key to understanding the co-evolution between galaxies and their central supermassive black holes (SMBHs), with BH mass $6 < \log_{10} (M_{\rm BH}/{\rm M}_\odot) < 9$ \citep[see e.g.,][]{dimatteo, treister}. Dual AGN represent the direct precursors of gravitationally bound binary SMBHs at parsec‑scale separations \citep[][]{volonteri2022}. These systems may eventually coalesce and produce gravitational‑wave emission that can be detected with pulsar timing arrays (PTAs; e.g., \citealt{agazie2023,ePTA}) and, at higher frequencies, with the forthcoming Laser Interferometer Space Antenna (LISA; \citealt{amaro,Colpi2024}) and Lunar Gravitational Wave Antenna (LGWA; \citealt{lgwa}). One of the most important open questions so far concerns the way through which AGN are triggered inside their host galaxy. Galaxy mergers are thought to cause the fuelling of gas onto the two nuclei, enhancing the accretion onto the SMBHs, which become active \citep{dimatteo, treister} and increase their average intrinsic luminosity as the separation between the nuclei decreases \citep{silverman, kocevski,koss,satyapal,hou,derosa}. When the merging galaxies are very massive, the inflows can even lead to the formation itself of a direct-collapse SMBH \citep{begelman, mayer2010, mayer2024}. Even though dual AGN are a fundamental ingredient in understanding the physics behind the co-evolution between galaxies and their central SMBHs, detecting them is still highly challenging, due to the spatial resolution of current instruments (which does not allow us to resolve  closer systems as redshift increases). A further limitation lies in the proper identification of active nuclei, which can often be obscured by a significant amount of cold gas (see \citealt{derosa2019} for a  review on theory and observations of dual AGN). In particular, in the scenario of a merger-driven AGN triggering, the fuelling of gas occurring during the merger could strongly obscure the nuclei, especially in the final stages of the merger \citep{hopkins}, thus decreasing the probability of detecting such systems when using ultraviolet (UV)/optical selections. Indeed, samples of dual AGN were found to suffer from higher obscuration than isolated ones \citep{derosa3, kocevski2, satyapal2, ricci2, derosa2, pfeifle, ricci, guainazzi, derosa}. This is also expected from simulations \citep{blecha2018}, although it is not yet clear how the column density evolves through the merger phases. This leads to a still incomplete census of dual-AGN systems, highlighting the need for complementary identification techniques across different wavebands \citep{comeford, fu, foord, mannucci}. These include imaging (for spatially resolved systems), spectroscopic diagnostics such as double-peaked [OIII] emission lines (for spatially un-resolved systems, \citealt{wang, smith1, ge, kim}), and/or combined methods such as integral field spectroscopy (e.g., \citealt{lena2015, mannucci, scialpi2024}), that combines spectroscopy with high-resolution imaging to confirm dual nuclei \citep[see][for a detailed description of dual-AGN identification techniques]{derosa2019}. Given these reasons, it is still difficult to make statistically significant studies of these sources (especially at the lower separations, i.e., $r_{\rm p}<30\unit{kpc}$), which are needed to compare the observed dual-AGN properties with simulations. So far, dual-AGN candidates have been identified primarily via optical observations (most of them through optical spectroscopy), but only hundreds of pairs have been confirmed to be real dual AGN through multi-wavelength studies, while the majority of them (some thousands) are still only candidate dual systems \citep{pfeifle2}. X-rays are so far one of the most efficient ways to detect AGN and, hence, dual AGN \citep{padovani}, since they allow us to pierce through and observe the most internal regions of the nucleus, where the primary AGN emission arises. Moreover, even though high absorption can affect the $2$--$10\unit{keV}$ X-ray emission, both intrinsic luminosity and $N_{\rm H}$ can be robustly recovered (up to obscuring column densities $N_{\rm H}<10^{24}\unit{cm^{-2}}$).

With the aim of investigating the properties of a homogeneous sample of AGN in pairs and the role of galaxy mergers in triggering the AGN activity, we present the study of an X-ray selected sample of AGN in pairs up to redshift $z\sim0.1$ (with projected spatial separation $r_{\rm p}\sim1$--$100\unit{kpc}$). Our work is organised as follows: in Section~\ref{sec.2}, we describe the sample selection; in Section~\ref{sec.3}, we present the X-ray data reduction and X-ray and optical spectral analysis of the AGN in our samples; our results are reported in Section~\ref{sec.4}; finally, in Section~\ref{sec.5}, we discuss our findings and provide a summary of our main results. We adopt a concordance cosmology with $H_0=70\unit{km\ s^{-1}\ Mpc^{-1}}$, $\Omega_\Lambda=0.7$, and $\Omega_{\rm M}=0.3$. The errors and upper/lower limits correspond to the 90\% confidence level, unless noted otherwise.

\section{Sample selection}\label{sec.2}

Dual AGN in X-rays have been studied by \citet{derosa}, who investigated a sample of optically selected narrow-line and broad-line AGN candidates in pairs up to $z\sim0.1$ from \citet{liu}, using XMM-Newton \citep{schartel} and Chandra \citep{weisskopf} data. AGN from \citet{liu} were selected using the Sloan Digital Sky Survey (SDSS; \citealt{york2000}). To confirm the AGN nature, \citet{derosa} combined multi-wavelength diagnostics using information from optical (BPT diagrams; \citealt{kauffman}), X-ray ($2$--$10\unit{keV}$ luminosity; \citealt{gonzalez}), and mid-infrared (mid-IR; using the ratio between X-ray and mid-IR fluxes versus hardness ratio, HR; \citealt{severgnini}) bands. Our aim is to enlarge this sample (composed of 34 AGN, 17 pairs) by searching for new AGN pairs using an X-ray-based selection, where the pair could be composed either of two AGN or of one AGN with a non-AGN galaxy companion. We selected our AGN sample by cross-correlating the 4XMM-Newton catalogue (DR14; \citealt{webb}) and the Chandra Source Catalog 2 \citep[CSC2;][]{evans2, evans} with the Millions of Optical Radio/X-Ray (MORX, v2)  Association catalog \citep{flesch}. The latter is an extensive catalogue of 3,115,575 optical sources associated with X-ray and radio sources (which can be AGN, type-I and II Seyferts, quasars, BL Lac objects, cataclysmic variable stars, galaxies, lensed quasars, and stars) detected with XMM-Newton, Chandra, Swift, ROSAT, VLASS, LoTSS, RACS, FIRST, NVSS, and SUMSS \citep{truemper, becker, mauch, woniak, burrows, shimwell, mcconnell, lacy}. An optical reference catalogue such as MORX is necessary to obtain the redshift, which in MORX is provided either as a spectroscopic or photometric measurement from the literature. For all sources selected in our sample, the redshift is provided by spectroscopic measurements. For this work, to have sufficiently high signal-to-noise-ratio X-ray data of AGN pairs, we limited our selection to $z < 0.1$ (for a total of 12,391 AGN candidates and 95,032 galaxies, as classified in MORX from the literature).

\subsection{AGN selection}\label{sub.2.1}

\begin{figure}
    \centering
    \includegraphics[width=0.8\linewidth]{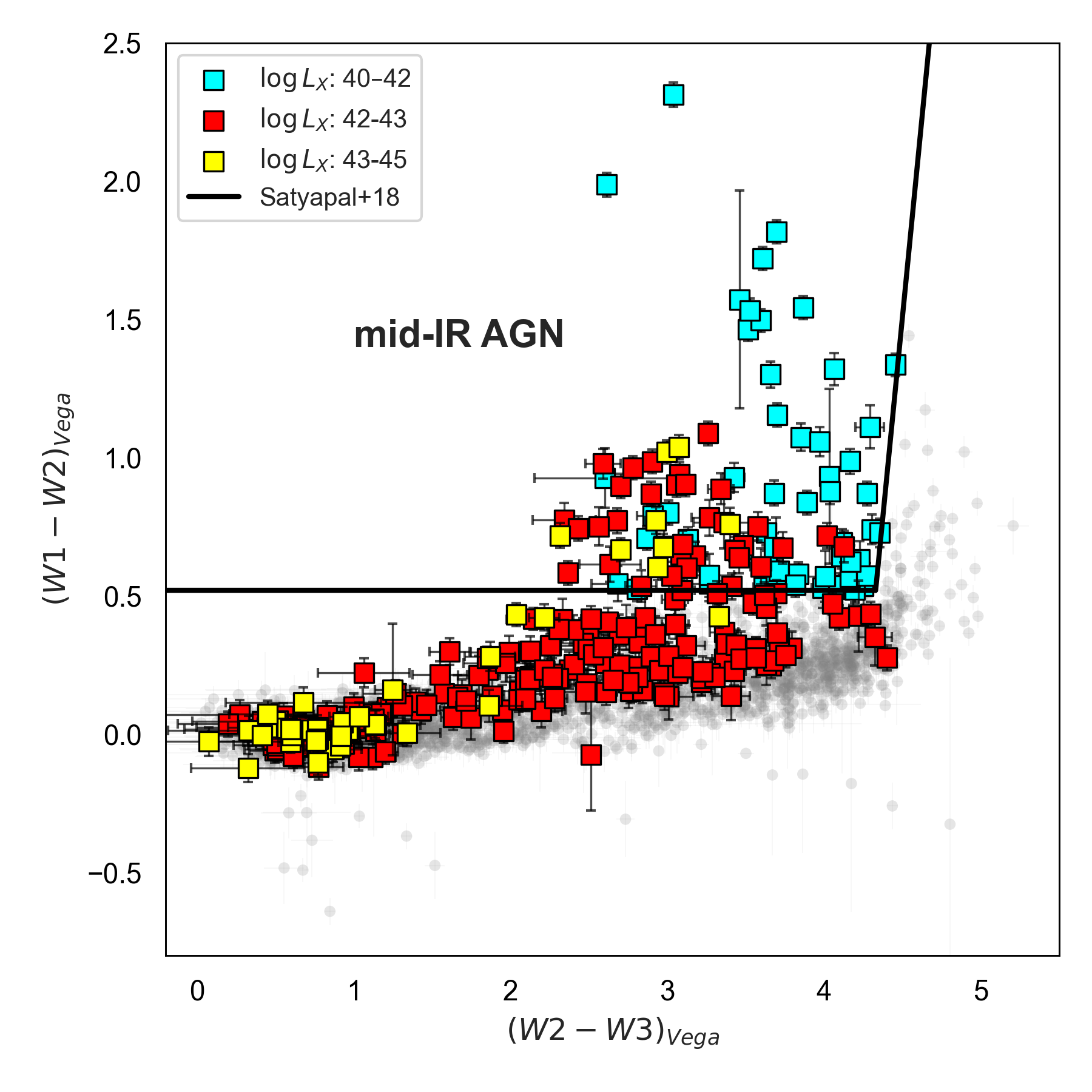}
    \caption{{\footnotesize WISE $W1$--$W2$ versus $W2$--$W3$ colour diagram used to select AGN amongst galaxies in MORX. The region identified by the black solid lines is the AGN region according to \citet{satyapal2018}. Squared/faded-circled points represent the galaxies that passed/did not pass the AGN test described in Section~\ref{sub.2.1}. Cyan, red, and yellow squares represent sources having $\log_{10} (L(2$--$10\unit{keV})\ /\ \unit{erg\ s^{-1}})=40$--$42,\ 42$--$43,\ {\rm and}\, 43$--$45$, respectively.}}
    \label{fig1}
\end{figure}

For the sources classified as AGN, Seyferts, and quasars in MORX, we cross-matched the latter catalogue with the XMM-Newton and Chandra source catalogues, using a $5^{\prime\prime}$ radius for the match position. This radius was adopted to account for the nominal astrometric uncertainties of both observatories for weak off-axis detections. Thus, by construction, all AGN in our sample are X-ray sources. From now on, we refer to this sample as the one obtained with XMM-Newton+Chandra. For our selection, we imposed thresholds on the X-ray fluxes of $F(2$--$12\unit{keV})>10^{-15}\unit{erg\ s^{-1}\ cm^{-2}}$ and $F(0.5$--$7\unit{keV})>10^{-16}\unit{erg\ s^{-1}\ cm^{-2}}$ for XMM-Newton and Chandra, respectively, given their sensitivity curves \citep{watson, evans2, evans}. We additionally imposed a $2$--$10\unit{keV}$ luminosity threshold of $L(2$--$10\unit{keV})>10^{40}\unit{erg\ s^{-1}}$. This threshold was chosen to avoid possible contamination from X-ray binaries in the host galaxy, considering that these sources are usually observed to have $2$--$10\unit{keV}$ luminosities up to $\sim$$10^{38}$--$10^{39}\unit{erg\ s^{-1}}$ \citep{grimm2002}. Moreover, previous studies showed that AGN usually have a higher $2$--$10\unit{keV}$ luminosity with respect to non-AGN sources \citep{gonzalez}. This led us to 1,737 X-ray-selected AGN candidates. In Figure~\ref{fig0}, we report all steps performed to select our final samples. For the sources classified as narrow-line AGN in MORX and with an optical spectrum available from the SDSS DR16 (\citealt{sdssdr16}, 428 sources), we also verified if they are classified as AGN in the BPT diagnostic diagram \citep{kauffman}. We found that four sources fall within the star-forming region, so we excluded them from the AGN sample. After this step, the effective number of AGN selected through optical and X-ray criteria  is 1,733. In order to recover possible “hidden” AGN amongst the objects classified as galaxy in MORX, we cross-matched the galaxy-classified sample with XMM-Newton and Chandra catalogues (again, using a $5^{\prime\prime}$ radius for the cross-match and only considering the sources fulfilling our X-ray criteria on flux and luminosity for AGN). We found 2,875 X-ray galaxies satisfying these requirements. We further applied two additional AGN selection criteria, requiring each source to satisfy at least one of them:\\
1. X-ray criterion: $L(2$--$10\unit{keV})>10^{42}\unit{erg\ s^{-1}}$. This is an additional requirement on the luminosity, and selects the brightest sources, which are therefore likely AGN (as also previously mentioned, see \citealt{gonzalez}). We stress that these sources are not optically classified as AGN in MORX. Consequently, when relying solely on X‑ray luminosity for AGN identification, a more conservative threshold is required.\\
2. Mid-IR criterion: mid-IR colours have been shown to be able to identify luminous AGN that are hidden in the optical band \citep{stern, mateos, secrest, andonie2025}. The main difficulty arises when the emission of the host galaxy is comparable to the emission of the AGN (as for bright starburst galaxies). Previous studies using Wide-field Infrared Survey Explorer (WISE; \citealt{wise}) colours revealed that a mid-IR diagnostic based on WISE $W1$--$W2$ colour (where $W1$ and $W2$ are the $3.4\unit{\mu m}$ and $4.6\unit{\mu m}$ magnitudes, respectively) can be used to identify AGN emission. For example, \citet{stern} identified a threshold based on $W1$--$W2>0.8$, which selects the most luminous AGN, whereas \citet{satyapal2014} use $W1$--$W2>0.52$ (which is more inclusive, but more contaminated by the presence of strong starburst galaxies). To exclude strong starburst galaxies contamination in the mid-IR band, \citet{satyapal2018} performed simulations combining two different WISE colours ($W1$--$W2$ versus $W2$--$W3$, where $W3$ is the WISE $11.6\unit{\mu m}$ magnitude), finding that, unlike starburst galaxies, AGN tend to occupy a precise region in the $W1$--$W2$ versus $W2$--$W3$ colour diagram (see Figure~\ref{fig1}). We then chose the latter criterion  to select AGN amongst the X-ray galaxies.\\
\noindent To summarize, an X-ray detected galaxy was considered as an AGN when: (a) it had an observed X-ray luminosity above $10^{42}\unit{erg\ s^{-1}}$ or (b) it had an observed X-ray luminosity above $10^{40}\unit{erg\ s^{-1}}$ and satisfied the mid-IR criterion from \citet{satyapal2018}. In this way, we found that 300 galaxies satisfied the above criteria (see Figure~\ref{fig1}) and we classified them as AGN (whose number thus increases from 1,733 to 2,033, see Figure~\ref{fig0}). Through this selection, we recovered, as expected, all the AGN from \citet{derosa} but four. These four AGN were not selected using our criteria for two possible reasons: 1) the source is not included in the MORX catalogue (J094554.4+423818, J103855.9+392157), or 2) it showed an X-ray  luminosity below the threshold we imposed (J015235.2-083233, J163103.4+395015). However, since all the four previous sources resulted as AGN in \citet{derosa} through their multi-wavelength diagnostics, we included them in our sample. After all these steps, we had a total of 2,037 X-ray AGN and 94,732 inactive galaxies.

\subsection{Pairs selection}

Dual AGN and AGN-galaxy pairs samples were selected by computing the angular separation between the nuclei, converted into a projected spatial separation using the redshift, and imposing a maximum $r_{\rm p}$ of $100\unit{kpc}$ and a velocity offset $\Delta v<300\unit{km\ s^{-1}}$ ($\Delta z<10^{-3}$). This criterion yields a total of  80 X-ray AGN in pairs (40 pairs), and 46 of them (23 pairs) appeared to
be newly identified \citep{hou, derosa}. The number of AGN-galaxy pairs is 122. We also identified a control sample of inactive galaxy pairs to compare statistical properties of AGN versus non-AGN pairs. Using the same thresholds for $r_{\rm p}$ and $\Delta z$, we found 2,930 inactive galaxies in pairs.

\subsection{Final samples}

\begin{figure}[t]
    \centering
    \includegraphics[width=\linewidth]{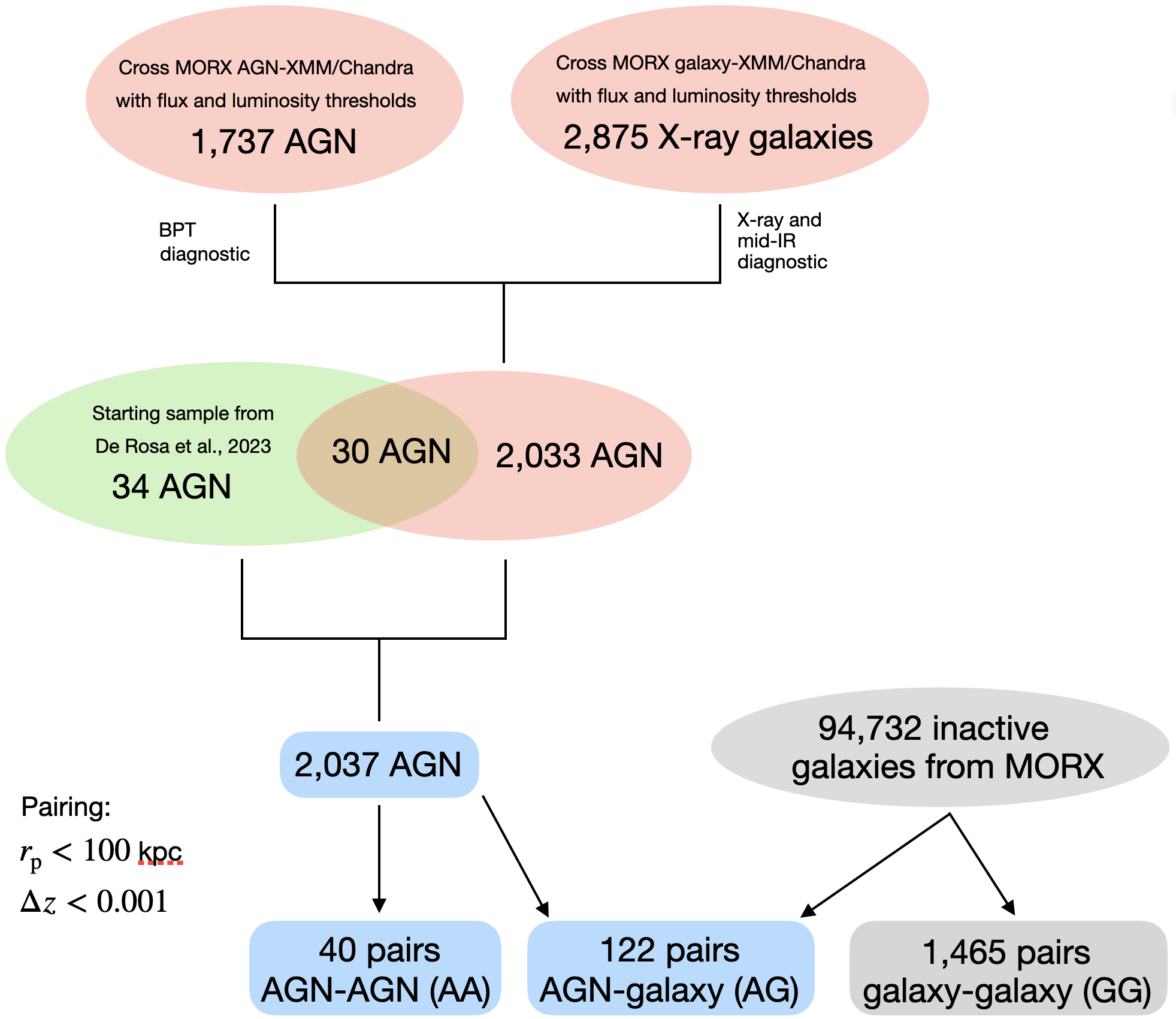}
    \caption{{\footnotesize Steps of our sample selection and resulting final numbers of AGN and galaxies in pairs.}}
    \label{fig0}
\end{figure}

We summarize the selection of AGN-AGN, AGN-galaxy, galaxy-galaxy, and isolated AGN samples as follows (see Figure~\ref{fig0}):

\begin{itemize}
    \item {AGN-AGN sample (AA)}: 80 AGN (40 pairs).
    \item AGN-galaxy sample (AG): 122 AGN (122 pairs).
    \item Total AGN pairs sample (TS): 202 AGN (162 pairs). We used this sample when referring to the total number of AGN in pairs  ({AA}+AG).
    \item Galaxy-galaxy sample (GG): 2,930 galaxies (1,465 pairs). This is the control sample of non-AGN galaxy-galaxy pairs.
    \item Parent sample (PS): this is the sample of isolated AGN, obtained as the difference between the total number of X-ray AGN in MORX and the number of AGN found in pairs, for a total of 1,835 isolated AGN.
\end{itemize}

\begin{figure}
    \centering
    \includegraphics[width=0.8\linewidth]{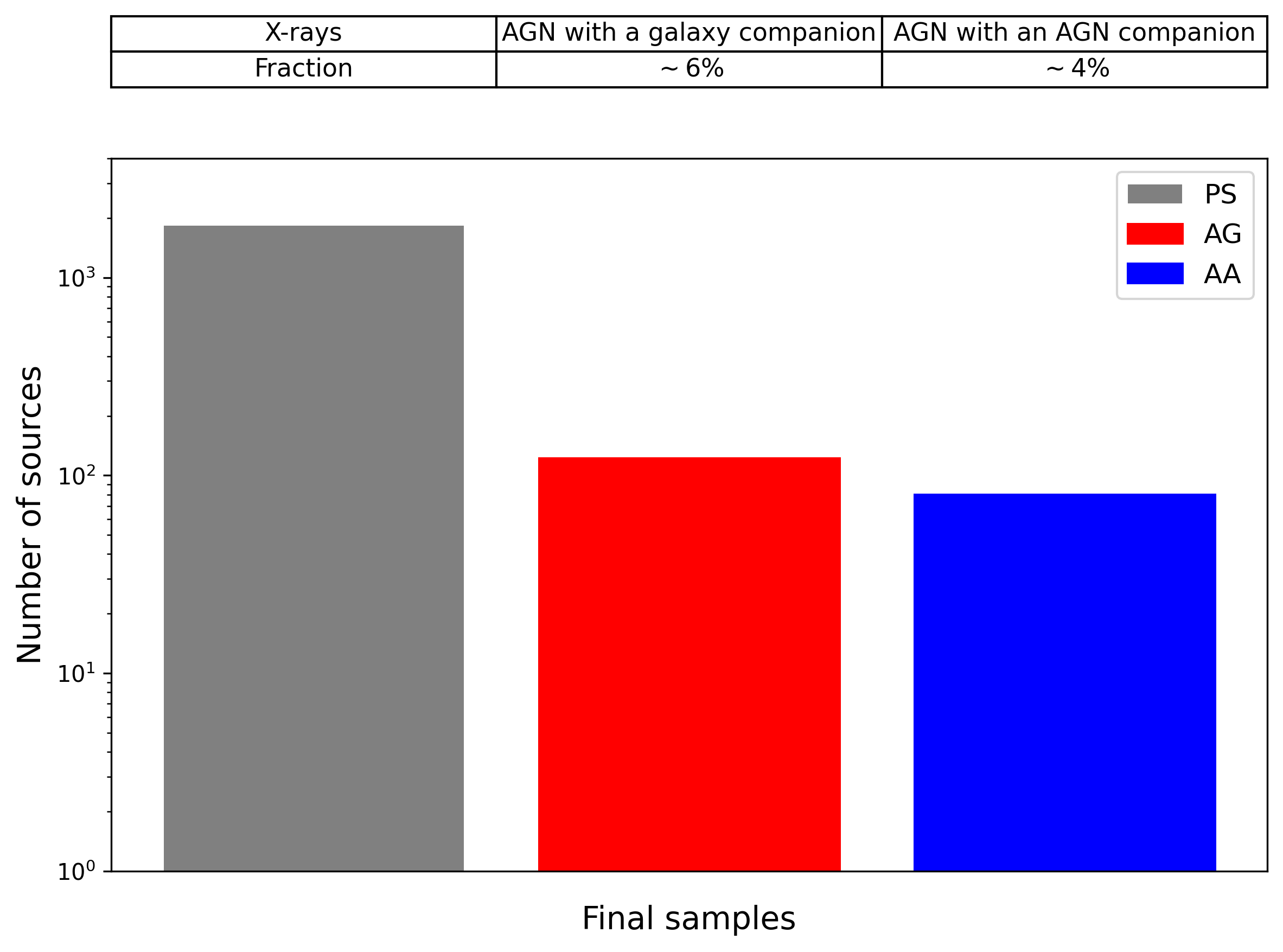}
    \caption{{\footnotesize Results from our sample selection: the upper table shows the fractions of AGN in pairs (with respect to the total number of AGN) considering all AGN with a galaxy companion and all AGN with an AGN companion. The plot below the table shows the number of isolated AGN (PS), AGN with a galaxy companion (AG), and AGN with an AGN companion ({AA}) in grey, red, and blue, respectively.}}
    \label{fig2}
\end{figure}

The resulting numbers of AGN in the {AA}, AG, and PS are shown in Figure~\ref{fig2}. On average, we found that the fraction of AGN with a companion (either another AGN or a galaxy) is $\sim$10\%: the fraction of AGN in AGN-galaxy pairs and in AGN-AGN pairs is $\sim$6\% and $\sim$4\%, respectively (see the upper table in Figure~\ref{fig2}). We defined three different merger stages depending on the separation between the nuclei: late ($r_{\rm p}<30\unit{kpc}$), middle  ($30\unit{kpc}<r_{\rm p}<60\unit{kpc}$), and early state of merger ($r_{\rm p}>60\unit{kpc}$)\footnote{Note that this is a simplified distinction, since we are not considering the complete merger history of the two galaxies.}. The merger stages were chosen

\begin{figure}
    \centering
    \includegraphics[width=0.8\linewidth]{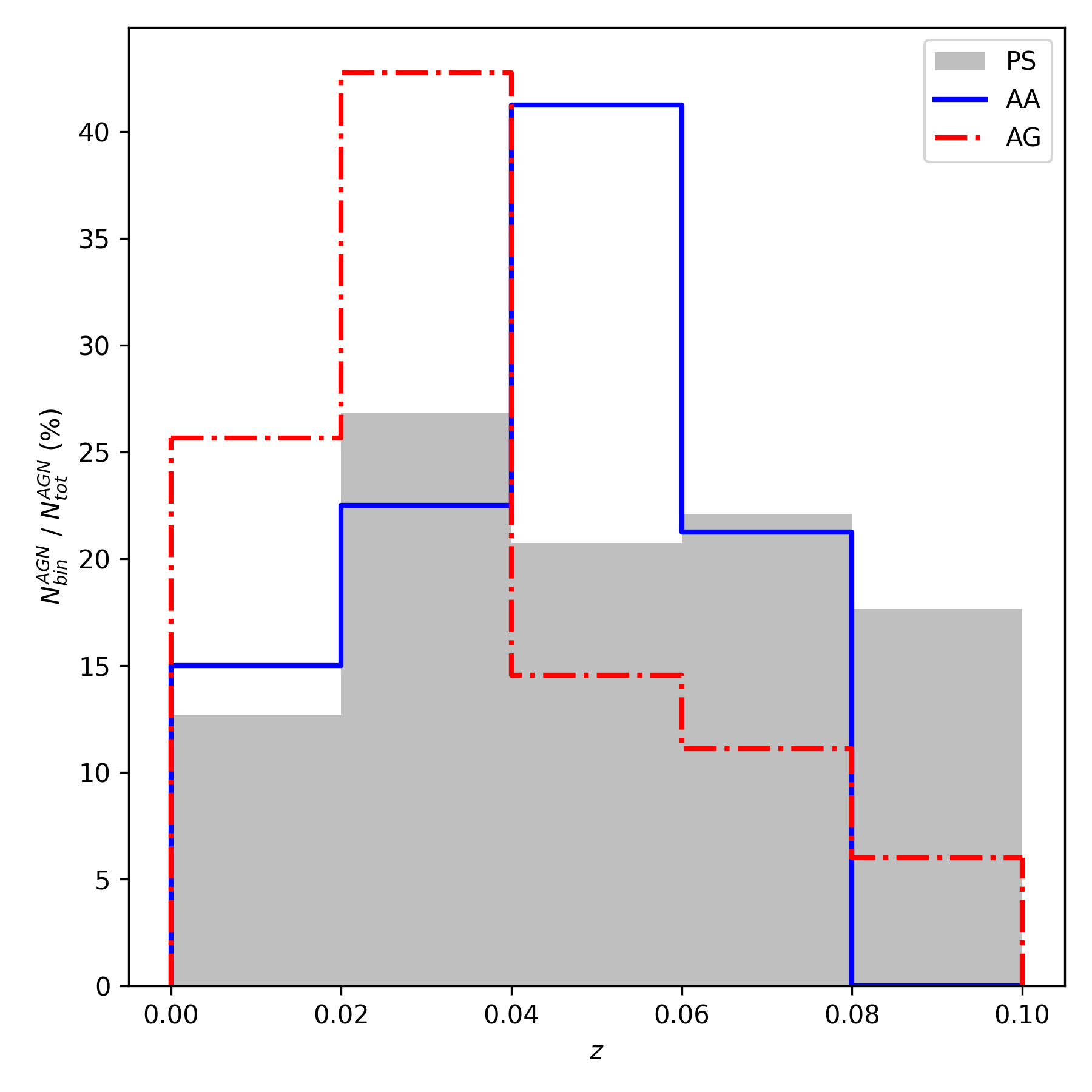}
    \caption{{\footnotesize Redshift distribution for the AGN in the {AA} (blue histogram), AG (dashed red histogram), and PS (shaded grey histogram). The fraction is computed as the number of AGN in the given redshift bin, over the total number of AGN in the given sample.}}
    \label{fig2_1}
\end{figure}

based on previous observational studies (e.g., \citealt{koss, derosa}) and simulations (e.g., \citealt{capelo2015}) on dual AGN, where it was found that the galaxy merger becomes relevant for AGN properties when $r_\mathrm{p}\lesssim30\unit{kpc}$. Amongst the pairs in the TS, 94 of them ($\sim$47\%) appear to be in a late state of merger, and most of them ($\sim$70\%) are pairs selected with Chandra, as we expect from its better angular resolution. The average redshift value for AGN in the {AA}, AG, and PS is $\sim$0.04, $\sim$0.03, and $\sim$0.05, respectively (the redshift distributions are shown in Figure~\ref{fig2_1} as blue, red, and grey histograms for the {AA}, AG, and PS, respectively).

\section{Data reduction and spectral analysis}\label{sec.3}

Our study included a combined X‑ray and optical spectral analysis of the {AA} pairs, complemented by an X‑ray analysis of the AG pairs. For the AGN in the PS, we also derived the relevant X‑ray properties through hardness‑ratio estimates. {The results of the X-ray and optical spectral analysis are shown in Table \ref{tab1} and \ref{tab3}, respectively.}

\subsection{X-ray}\label{sub.3.1}

For AGN in the {AA}, XMM-Newton (Chandra) X-ray spectra were extracted for AGN in pairs with at least $\sim$50 ($\sim$10) counts in the $0.2$--$12\unit{keV}$ ($2$--$8\unit{keV}$) EPIC PN (ACIS-S) band. For AGN in the AG, given the larger number of sources, we used the spectra available from the XMM-Newton and Chandra archives. We had a total of 52/80 available X-ray spectra of dual AGN (XMM-Newton+Chandra) and 71/122 (XMM-Newton+Chandra) X-ray spectra of AGN in the AG. Amongst dual AGN, 44 (22 pairs) are AGN for which the spectrum was extracted for both nuclei of the pair. For the AGN studied in \citet{derosa}, we refer to  results derived from their spectral analysis. We used the Science Analysis System (SAS) software V17 \citep{sas} for the XMM-Newton data reduction and the Chandra Interactive Analysis of Observations (CIAO) software V4.16 \citep{fruscione} for the Chandra data reduction. We applied proper cuts above $10\unit{keV}$ to exclude the flaring particle background for XMM-Newton. The spectra were extracted from regions with an aperture radius of at least $\sim$$20''$ and $\sim$$1''$ for XMM-Newton and Chandra, respectively, in order to include at least $80\%$ of the source counts at $\sim$$1.5\unit{keV}$ for both instruments, given their Encircled Energy Fraction. Source and background regions were extracted using the same radius, where the background region was selected to avoid contamination from the source. All  spectra were analysed using the software Xspec V12.14 \citep{arnaud} using a $\chi^2$ statistic when spectra had at least $100$ ({$60$}) net counts in the $0.2$--$12\unit{keV}$ ({$0.5$}--$8\unit{keV}$) band for XMM-Newton (Chandra), and using a Cash statistic \citep{cash, wachter} when net counts were lower. {When using} {$\chi^2$} {statistic, we ensured to have at least 20 (15) counts in each bin for XMM-Newton (Chandra).} We assumed as baseline model (BLM) for the X-ray emission of our AGN the following:\\
- a primary X-ray emission of the AGN corona in the $2$--$10\unit{keV}$ band due to the comptonization of photons from the accretion disc of the SMBH \citep{haardt, haardt2, ursini2023, gianolli2023}. This component can be described by a power law;\\
- an absorption component from neutral gas with a column density $N_{\rm H}$, which obscures the primary AGN emission;\\
- a soft X-ray emission, below $\sim$$2\unit{keV}$, approximable with a power law in the majority of our spectra. For obscured AGN, this component could be either attributed to the emission from the narrow-line region (see, e.g., \citealt{bianchi}), to the scattering of the primary emission of the AGN \citep{ueda2007}, or to the galaxy star formation \citep[][]{ranalli2003}; for un-obscured AGN, this component could be attributed to reflection processes from the inner parts of the accretion disc and/or to the absorption from an outflowing warm gas \citep[][]{miniuttu2004, gierlinski, chen2025};\\
- an absorption due to the Milky Way column density along the line of sight \citep{hi4pi}.\\
\noindent Therefore, our BLM in Xspec is $TBabs\times(zTBabs\times powerlaw^H+powerlaw^S)$, which has the form\\
\\
\noindent $f(E)=e^{-N^{\rm Gal}_{\rm H}\sigma}\left(e^{-N_{\rm H}\sigma}K^HE^{-\Gamma^H}+K^SE^{-\Gamma^S}\right)$\,,\\
\noindent where $N^{\rm Gal}_{\rm H}$ is the hydrogen column density of the Milky Way, $\sigma$ is the photoelectric absorption cross-section \citep{balucinska}, $N_{\rm H}$ is the hydrogen column density obscuring the AGN, $K$ is the normalisation of the power law at $1\unit{keV}$, and $\Gamma$ is the photon index of the power law. The superscripts $H$ and $S$ refer to the hard (primary, $2$--$10\unit{keV}$) and soft ($<2\unit{keV}$) component, respectively. For 18 spectra, we only needed one power law to model the data. In 10 sources, the spectrum showed evidence of reflection features believed to be produced by the reprocessing of the primary AGN emission from the molecular torus \citep{antonucci, jaffe, meisenheimer, raban}, the broad-line and narrow-line regions (\citealt{bianchi2008, ponti}), or the accretion disc \citep{george, matt91}. Such features usually arise as emission line components, such as the Fe\ K$\alpha$ line at $6.4\unit{keV}$ and a Compton hump above $\sim$$10\unit{keV}$, peaking at $\sim$$30\unit{keV}$. In such cases, we added the Fe\ K$\alpha$ line as a Gaussian profile and used the $pexrav$ model in Xspec to reproduce only the reflection component above $\sim$$10\unit{keV}$. The latter is an exponentially cut-off power-law spectrum reflected from a neutral material \citep{magdziarz}. When a simple power-law model was not sufficient to reproduce the soft X-ray emission, we used $Mekal$ model as an additive component. The latter is an emission spectrum from a hot diffuse gas that includes several emission lines \citep{mewe, mewe2, liedahl}. When performing the fits, we typically allowed $\Gamma^H$ and $N_{\rm H}$ to vary, while the value of $\Gamma^S$ was  usually fixed to that of $\Gamma^H$. When it was not possible to estimate simultaneously $\Gamma^H$ and $N_{\rm H}$, or their associated errors, we fixed the photon-index value to a typical 1.9, as observed in large populations of local AGN \citep{nandra, bianchi2,akylas}.

\subsection{Optical}\label{sub.3.2}

For the optical analysis, we focused on dual AGN in the {AA} in which both nuclei had an optical spectrum available from SDSS DR16: 46 AGN, on 23 out of the 40 pairs of the GS. Our choice is motivated by the goal of comparing the properties of the two BHs in each pair, as will be described in the results in Section~\ref{sec.4}. The optical spectra were analysed using the Penalized Pixel-Fitting (pPXF) method by \citet{cappellarippxf}, which allows to perform the fitting of stellar kinematics and gas optical emission lines. For the stellar population, we used the E-MILES templates \citep{emiles}. When performing the fits, the noise level was chosen in order to have $\chi^2/\unit{d.o.f.}\sim1$ for each spectrum, assuming a constant noise per pixel. For the fitting procedures, we separated the stellar models from the gas emission lines as different components. The emission line components include the narrow Balmer and [O$\rm III]5007\angstrom$ lines, which we used to estimate the bolometric luminosities of sources without an X-ray spectral analysis, as described in the following. Since all sources but one (J103853.29+392151.1\footnote{Our mass measurement for this source is consistent, within the uncertainties, with the one obtained in \cite{derosa2} using Single Epoch mass estimates.}) show only narrow emission lines, we used the stellar velocity dispersion ($\sigma^*$) found through the fitting of the stellar kinematic component to estimate the BH mass. To do this, we used the relation from \citet{mcconnel2013}:

\begin{equation}
    \resizebox{0.93\columnwidth}{!}{%
    $ \log_{10}{\left(\frac{M_{\rm BH}}{{\rm M}_\odot}\right)}=(8.32\pm0.05)+(5.64\pm0.32)\log_{10}{\left(\frac{\sigma^*}{200\unit{km\ s^{-1}}}\right)}\,. $}
\end{equation}

Using these mass estimates, we derived the corresponding \citet{Eddington} luminosity ($L_{\rm Edd}$) of each AGN in the 23 pairs with an optical spectrum. We then proceeded to estimate their bolometric luminosity ($L_{\rm bol}$), which is needed to compute the Eddington ratio ($\lambda_{\rm Edd}$, defined as $L_{\rm bol}/L_{\rm Edd}$). We assumed the bolometric correction factor from \citeauthor{duras2020} (\citeyear{duras2020}; which uses the AGN X-ray luminosity) for the AGN which also had an X-ray spectral analysis. When the AGN counts were too low to allow an X-ray spectral extraction and analysis (see Section~\ref{sub.3.1}), we used the bolometric luminosity estimated from \citet{netzer2009}, which requires the narrow H$\beta$ and [O$\rm III]5007\angstrom$ fluxes and the extinction-corrected H$\beta$ luminosity (described below). The observed fluxes and luminosity cited above were measured  fitting the gas emission lines with pPXF. Additionally, to correct the luminosity of H$\beta$, we estimated the dust reddening correction $E(B$--$V)$ from
\begin{figure}[t]
    \centering
    \includegraphics[width=0.8\linewidth]{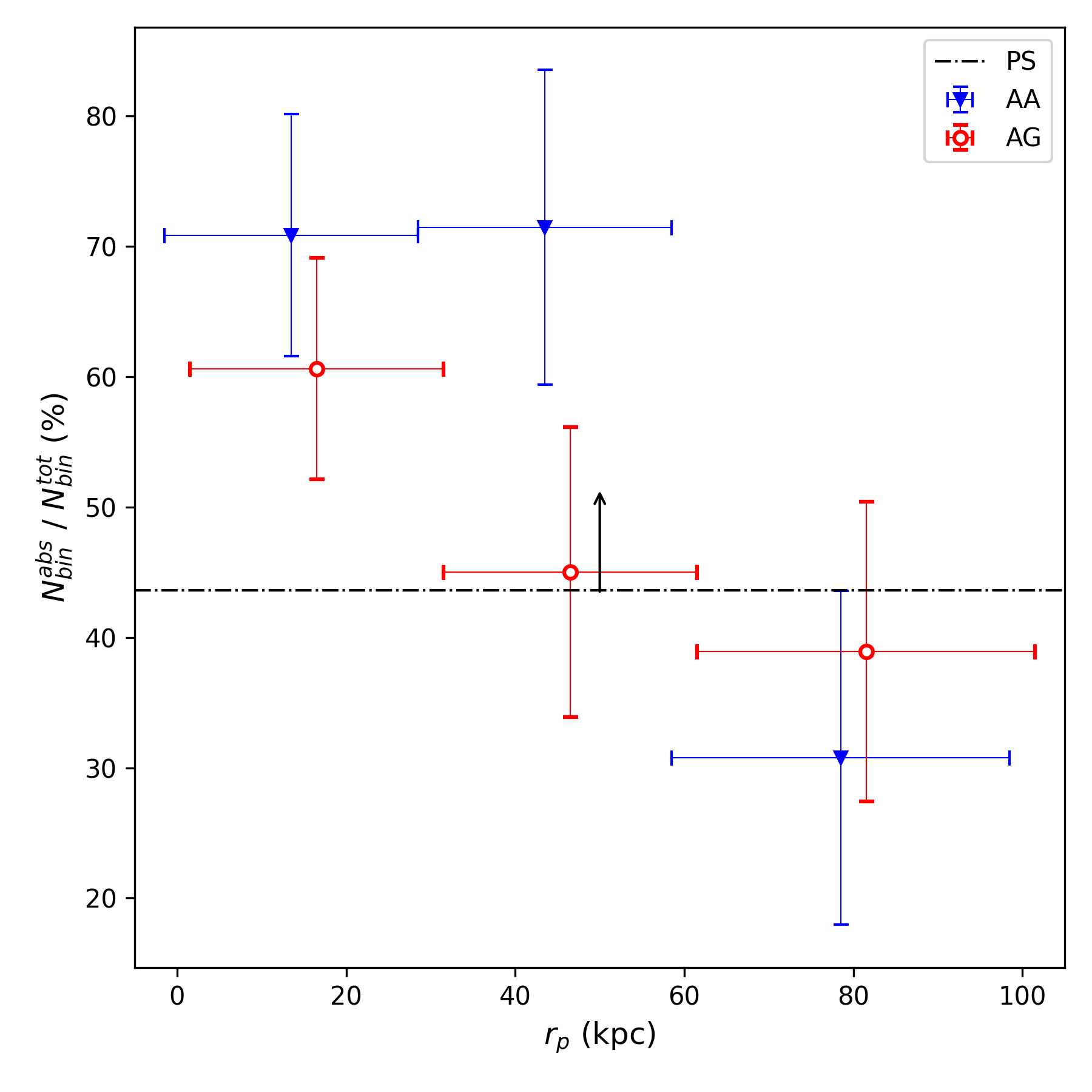}
    \caption{{\footnotesize Fraction of absorbed ($N_{\rm H}\geq10^{22}\unit{cm^{-2}}$) AGN in the {AA} (blue data) and AG (red data) in bins of $0$--$30\unit{kpc}$ (late state of merger), $30$--$60\unit{kpc}$ (middle state of merger), and $60$--$100\unit{kpc}$ (early state of merger). Fractions are computed as $N^{\rm abs}_{\rm bin}/N^{\rm tot}_{\rm bin}$, where $N^{\rm abs}_{\rm bin}$ and $N^{\rm tot}_{\rm bin}$ are the numbers of AGN with $N_{\rm H}\geq10^{22}\unit{cm^{-2}}$ in the given bin and the total number of AGN in that bin, respectively. For AGN in the {AA} and AG, the fractions are computed using $N_\mathrm{H}$ from spectral analysis, while for AGN in the PS (dashed, horizontal line) the fraction is estimated as a lower limit through XMM-Newton HR3 and HR4.}}
    \label{fig3}
\end{figure}
the extinction law by \citet{cardelli2006}, assuming a dust screen, Case-B recombination at a gas temperature of $10^4\unit{K}$ \citep{osterbrock2006}. For one AGN without X-ray spectral analysis (J011832.8-000936), the SDSS spectrum shows a gap between $\sim$$6200\angstrom$ and $\sim$$7000\angstrom$, where the H$\alpha$ line falls. Thus, for this source, we corrected the H$\beta$ luminosity considering the mean $E(B$--$V)\sim0.3$ value found in our sample. For sources with both X-ray and optical information, the derived bolometric luminosities are consistent.

\section{Results}\label{sec.4}

In this section, we present the main results of our selection and analysis. In particular, we focused on the absorption properties of dual AGN in the {AA} and AG (Section~\ref{sub.4.3}) and their triggering mechanisms (e.g., luminosity and accretion properties; Section~\ref{sec.4.2}). All the uncertainties on fractions are computed using binomial errors. When the fractions were computed using values from spectral analysis (e.g., $N_{\rm H}$),  uncertainties are propagated considering the binomial error and the error derived from the spectral analysis.

\subsection{Absorption properties}\label{sub.4.3}

\subsubsection{Absorption properties from spectral analysis}

One of the main goals of our work is to study the absorption properties of AGN in pairs in the {AA} and AG, and to eventually distinguish them from those of isolated AGN in the PS. Thanks to our spectral analysis, we were able to measure the column density of 52 and 71 AGN of the {AA} and AG, respectively. We found a fraction of absorbed AGN ($N_{\rm H}\geq10^{22}\unit{cm^{-2}}$) of $61\pm7\%$ ($51\pm8\%$) in the {AA} (AG), while the fraction of Compton-thick (CT, $N_\mathrm{H}>10^{24}\unit{cm^{-2}}$) AGN is $8\pm4\%$ ($4\pm2\%$). Our classification of obscured and un-obscured AGN is robust against the presence of $N_{\rm H}$ upper and lower limits found through spectral analysis. Specifically, all lower limits are associated to  obscured sources, while  all upper limits correspond to un-obscured ones, ensuring that  the relative fractions remain unchanged. The fractions found for dual AGN are in good agreement with what was found in previous studies on dual AGN \citep{koss2011, ricci2, ricci, derosa}, and are generally higher than the fractions of Compton-thin AGN found for samples for isolated AGN using other surveys (e.g., $\sim$45\% using BAT, \citealt{krimm, ricci6}). The fraction of CT AGN in pairs found in our study (and similarly the $\sim$16\% reported by \citealt{derosa}) is lower than what has been found in some studies of isolated AGN. For example, \citet{boorman} studied a homogeneous sample of isolated AGN ($z<0.044$) using the NuSTAR \citep{harrison} local AGN $N_{\rm H}$ distribution survey (NULANDS), finding a fraction of $\sim$35\% of CT AGN. It is worth noticing that estimating such regimes of obscuration could be highly challenging when using instruments such as XMM-Newton and Chandra. In fact, when $N_{\rm H}\gtrsim10^{24}\unit{cm^{-2}}$, the primary component of the AGN is suppressed up to $10\unit{keV}$, which is the energy limit we can investigate with XMM-Newton and Chandra at $z < 0.1$. Thus, the fraction of CT AGN in pairs we found in this analysis should be regarded as a lower-limit. Finally, to investigate a possible evolution of the obscuration with merger stages in AGN  pairs, we estimated the fractions of obscured AGN (referring to those with a measured $N_{\rm H}$) in bins of projected spatial separations $r_{\rm p}$ (see Figure~\ref{fig3}). The fractions are computed as $N^{\rm abs}_{\rm bin}/N^{\rm tot}_{\rm bin}$, being $N^{\rm abs}_{\rm bin}$ and $N^{\rm tot}_{\rm bin}$ the number of absorbed AGN and the total number of AGN in the given bin, respectively. We chose bins of $<$$30\unit{kpc}$ (late state of merger), $30$--$60\unit{kpc}$ (middle state of merger) and $60$--$100\unit{kpc}$ (early state of merger). As a result, for AGN in the {AA} (blue data points in Figure~\ref{fig3}), we found that the fraction of absorbed AGN is higher at the closest separations ($0$--$30\unit{kpc}$ and $30$--$60\unit{kpc}$), where it reaches about $\sim$70\% for {AA} AGN in late state of merger. Conversely, in the early state of merger, the fraction of absorbed AGN is significantly lower, at about $\sim$30\%. Moreover, the fraction of obscured AGN in the late state of merger is significantly higher than the fraction (estimated through the HR) of obscured isolated AGN in the PS (although we stress that the latter fraction is still a lower limit, see Section~\ref{sub.4.1.2}).  Although for the AG the error bars prevent us from identifying a clear trend in the obscured fraction with separation (red data points in Figure~\ref{fig3}), there is marginal evidence for an increase in the obscured AGN fraction across the merger stages, albeit slightly lower than that observed for the dual-AGN sample. A similar result for AGN in pair with galaxies was found also in previous studies using X-ray data (e.g., \citealt{guainazzi}, where $N_\mathrm{H}$ was found to increase for decreasing $r_\mathrm{p}$). These findings suggest that mergers have an impact in the AGN obscuration, likely caused by the increased fuelling of gas within the two galaxies, as also expected from simulations (\citealt{hopkins,capelodotti2017,blecha2018,volonteri2022}; for a recent review, see \citealt{capelo2023}) and as also found in previous studies on dual AGN \citep{derosa3, kocevski2, ricci_bat, satyapal2, derosa2, pfeifle, guainazzi, ricci}.

\subsubsection{Absorption properties from the hardness ratio}\label{sub.4.1.2}

To estimate the fraction of Compton-thin AGN in the PS, and  to compare it with the fractions derived from the spectral analysis of the {AA} and AG, we carried out  XMM-Newton simulations in Xspec  to quantify the correlation between hardness ratios HR3 (energy bands $1$--$2\unit{keV}$ and $2$--$4.5\unit{keV}$) and HR4 (energy bands $2$--$4.5\unit{keV}$ and $4.5$--$10\unit{keV}$) and the column density $N_{\rm H}$. The simulations are described in detail in Appendix~\ref{sec:appendixb}. The HR3 values were used to distinguish between un-obscured and obscured AGN, while HR4  was used to find highly obscured AGN ($N_\mathrm{H}=10^{23}$--$10^{24}\unit{cm^{-2}}$) that could be missed using only ``soft'' HR3. Looking at simulations (described in Appendix~\ref{sec:appendixb}; see Figure~\ref{Figurea1}), we can consider sources having HR3$\,>0$ and HR4$\,>0.2$ as absorbed with $N_\mathrm{H}>10^{22}\unit{cm^{-2}}$ and as highly absorbed with $10^{23} \leq N_\mathrm{H}/\unit{cm^{-2}} \leq 10^{24}$, respectively. In the PS, 949 AGN are detected with XMM-Newton and have HR3 and HR4 available from the catalogue. Using the HR3 and HR4 thresholds mentioned above, we found that $\gtrsim$44\% of AGN in the PS are absorbed. This fraction (although it is a lower limit) is in agreement, within the error bars, with the fraction of obscured AGN in the early stage of merger found in the {AA} and AG through spectral analysis. Moreover, this result is also consistent with the fraction ($\sim$45\%) of Compton-thin isolated AGN found in BAT \citep{krimm, ricci6}, as previously mentioned. We stress that even combining HR3 and HR4 values does not allow us to reliably identify CT AGN. Thus, these fractions should be considered as a lower limit. However, the same bias in identifying CT AGN also arises in the spectral analysis if we use XMM-Newton and Chandra, as already discussed. This gives quite comparable results between spectral analysis and HR-based estimations.

\subsection{AGN triggering}\label{sec.4.2}

\subsubsection{Fraction of AGN in pairs}\label{sub.4.1}

In order to investigate the triggering mechanisms of AGN in pairs, we first look at the AGN fractions derived from our sample selection. As mentioned in Section~\ref{sec.2}, we found a fraction ($N^{\rm AGN}_{\rm pair}/N^{\rm AGN}_{\rm tot}$) of AGN in pairs (including both dual AGN and AGN-galaxy pairs) of $\sim$10\% of all the X-ray selected AGN below $100\unit{kpc}$ (4\% for late-state mergers, i.e., for separations below 30 kpc). Within this fraction, $\sim$4\% ($\sim$2\% for late-state mergers) are dual AGN  (see Figure~\ref{fig2}). These fractions are in agreement with previous studies of dual AGN in the X-ray band at low redshift (e.g., \citealt{he2023}). To investigate the connection between AGN activity and galaxy mergers, we quantified the excess of AGN pairs in the TS relative to the incidence of galaxy pairs in the GG. The excess was computed in bins of projected separation of width of $20\unit{kpc}$.
For each bin in $r_{\rm p}$, we defined the excess as ${f_{\rm AGN}}/{f_{\rm gal}}$, where $f_{\rm AGN}$ and $f_{\rm gal}$ represent the fractions of AGN and galaxies in pair in that separation bin, respectively. These fractions were calculated as
\[
f_{\rm AGN} = \frac{N^{\rm AGN,TS}_{\rm bin}}{N^{\rm AGN}_{\rm tot}}\,, \\
f_{\rm gal} = \frac{N^{\rm gal,GG}_{\rm bin}}{N^{\rm gal}_{\rm tot}}\,,
\]
\noindent with $N^{\rm AGN,TS}_{\rm bin}$ and $N^{\rm gal,GG}_{\rm bin}$ being the number of AGN and galaxies in pair within the corresponding bin of $r_{\rm p}$. {The excess was evaluated considering, for both $N^{AGN,TS}_{bin}$ and $N^{AGN}_{tot}$: the number of all AGN, the number of obscured AGN, and the number of un-obscured AGN. For the latter two, we considered the obscuration found through X-ray spectral analysis (for AGN pairs) and HR-based
\begin{figure}
    \centering
    \includegraphics[width=0.8\linewidth]{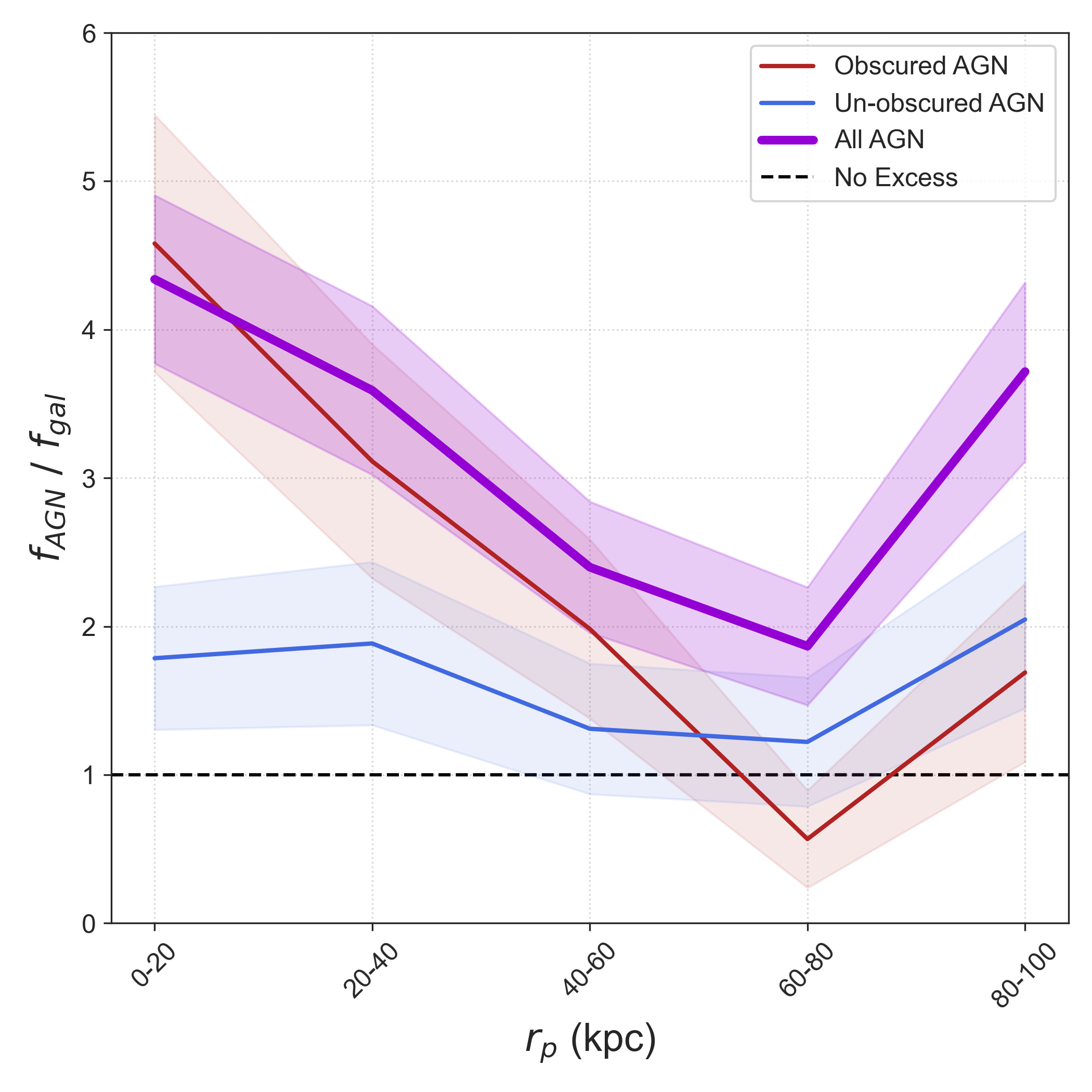}
    \caption{{\footnotesize Excess of AGN  pairs {in TS} with respect to galaxy pairs (GG) as a function of the projected spatial separation. The excess is evaluated as $f_{\rm AGN}/f_{\rm gal}$, being $f_{\rm AGN}$ the fraction of AGN in pairs in a given bin with respect to the total number of AGN (pairs+isolated), and $f_{\rm gal}$ the same as $f_{\rm AGN}$ but for galaxy pairs. {The purple line represents the excess for all AGN in TS, whereas the red and blue lines show the AGN excess only for obscured and un-obscured AGN (using the results from the X-ray spectral analysis), respectively.} The black, dashed horizontal line represents the no-excess case.}}
    \label{fig4}
\end{figure}
estimations (for isolated AGN), see Section~\ref{sub.4.3}.} The total {number for galaxies is {$N^{\rm gal}_{\rm tot}=94{,}732$}, so that {$f_{\rm gal}$} is constant (in the corresponding bin) for all three groups of AGN (total, obscured, and un-obscured). Instead, {$N^{\rm AGN}_{\rm tot}$} depends on the classification of AGN for which we evaluated the excess, and corresponds to {$2{,}037$} for all AGN, {$874$} for obscured AGN, and {$1{,}083$} for un-obscured AGN. All total numbers include both paired and isolated AGN (or galaxies). The AGN pairs excesses are shown in Figure~\ref{fig4}. As a result, we found an average excess of {$\sim3$} for all AGN (obscured+un-obscured) with respect to galaxies when looking at sources in pairs. That could mean that AGN are, on average, more common than inactive galaxies when in pairs. However, the excess for the total number of AGN in TS does not show a significant trend with separation (with Spearman {$\rho\sim-0.4$}, {$p$}-value {$\sim0.5$}, see the purple line in Figure \ref{fig4}). Interestingly, if we instead separate the excess for obscured and un-obscured AGN (see the red and blue lines in Figure \ref{fig4}, respectively), a significant increase of the excess with decreasing {$r_\mathrm{p}$} arises for obscured AGN (with Spearman {$\rho\sim-0.9$}, {$p$}-value {$\sim0.04$}). The same trend is not observed for un-obscured AGN. This suggests that AGN, when obscured, are more common in pairs with respect to galaxies, especially at the closest separations (as expected from simulations; \citealt{capelo2017}). These results are in good agreement with the higher fraction of obscured AGN at the closest separation found through X-ray spectral analysis (see Figure~\ref{fig3}), and show that galaxy mergers may have an important role in triggering AGN activity and may be responsible for a high obscuration of the nuclei. Similar results were also found in other studies on AGN in pairs \citep{gao,ellison2025} that revealed an increase of narrow-line and mid-IR-selected AGN excess with decreasing {$r_\mathrm{p}$} (with a peak right after the coalescence phase).}

\begin{figure*}
\centering
    \includegraphics[width=0.6\linewidth]{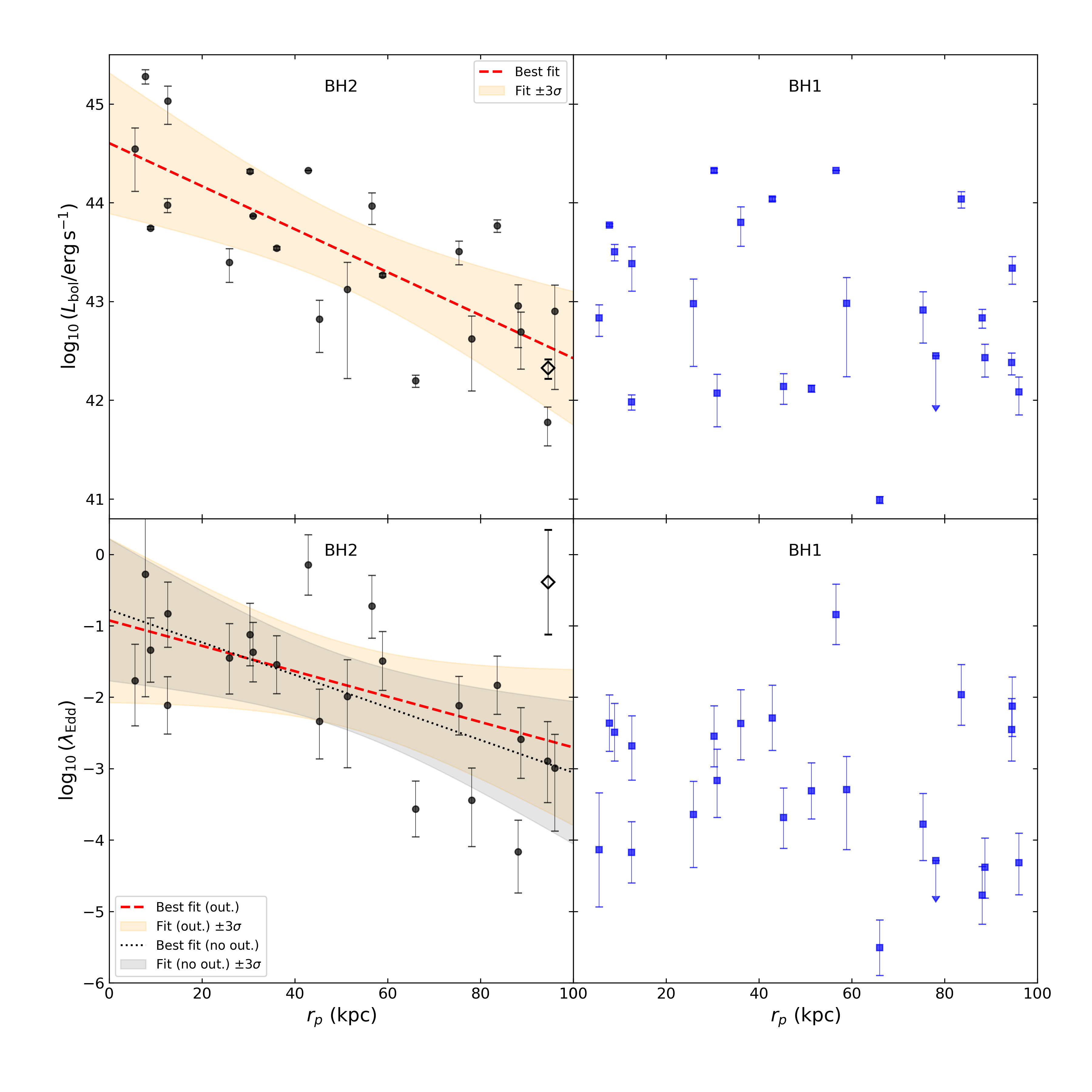}
    \caption{{\footnotesize Bolometric luminosity and Eddington ratio as a function of projected spatial separation. Upper panels (from left to right): bolometric luminosity of BH2 as a function of projected spatial separation. The red dashed line and the yellow shaded band represent the best-fit relation and its uncertainty region at $\pm3\sigma$; bolometric luminosity of BH1 as a function of projected spatial separation. Lower panels (from left to right): Eddington ratio of BH2 as a function of the projected spatial separation. The red dashed (black dotted) line and the yellow (grey) shaded area represent the best-fit relation and its uncertainty region at $\pm3\sigma$, including (excluding) IMBH J085125.8+393541; Eddington ratio of BH1 as a function of the projected spatial separation.}}
    \label{fig5}
\end{figure*}

\subsubsection{AGN luminosity and accretion}\label{sub.4.2}

\begin{figure*}[ht!] \centering \includegraphics[width=0.37\linewidth]{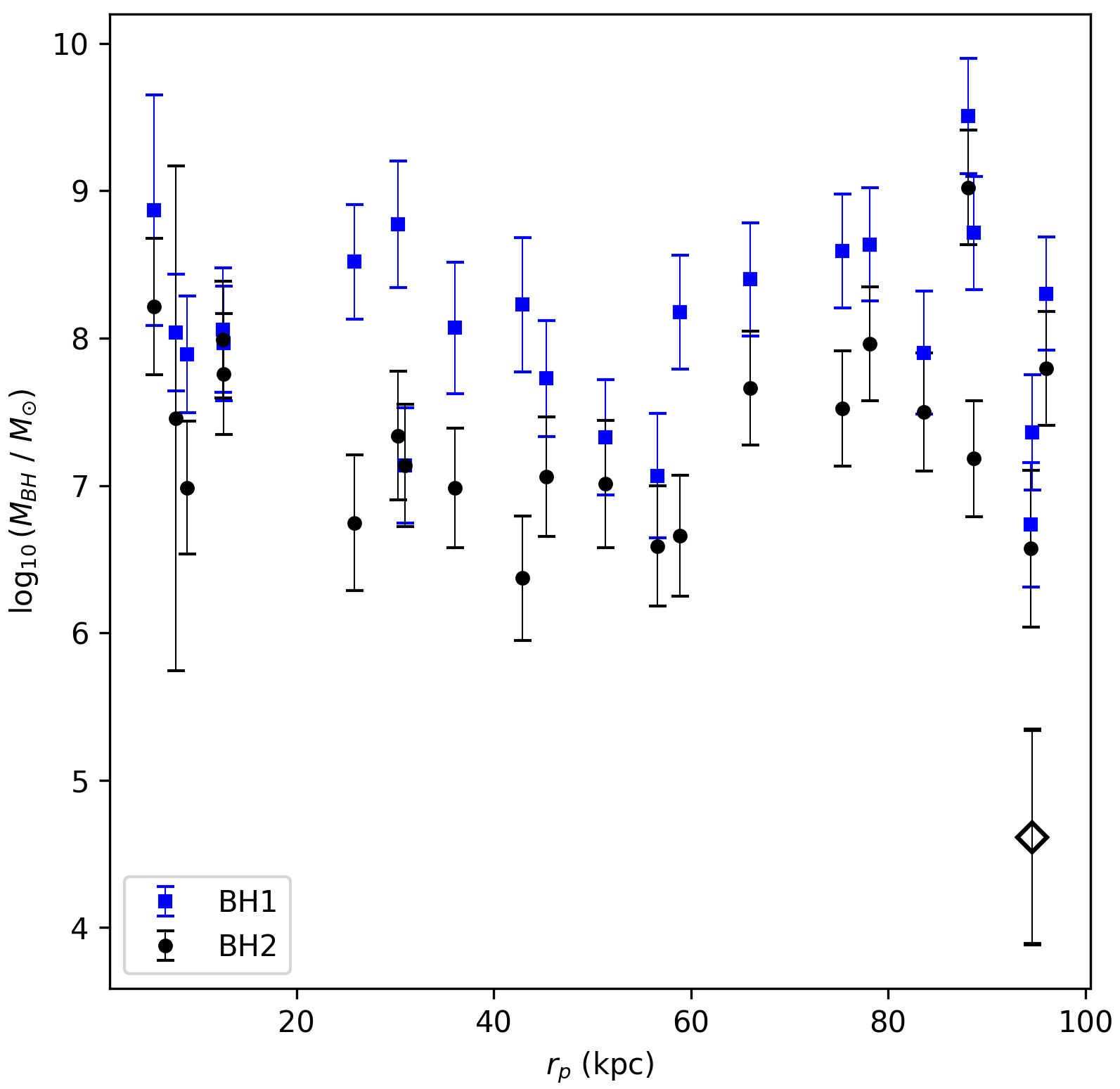} \hfill \includegraphics[width=0.45\linewidth]{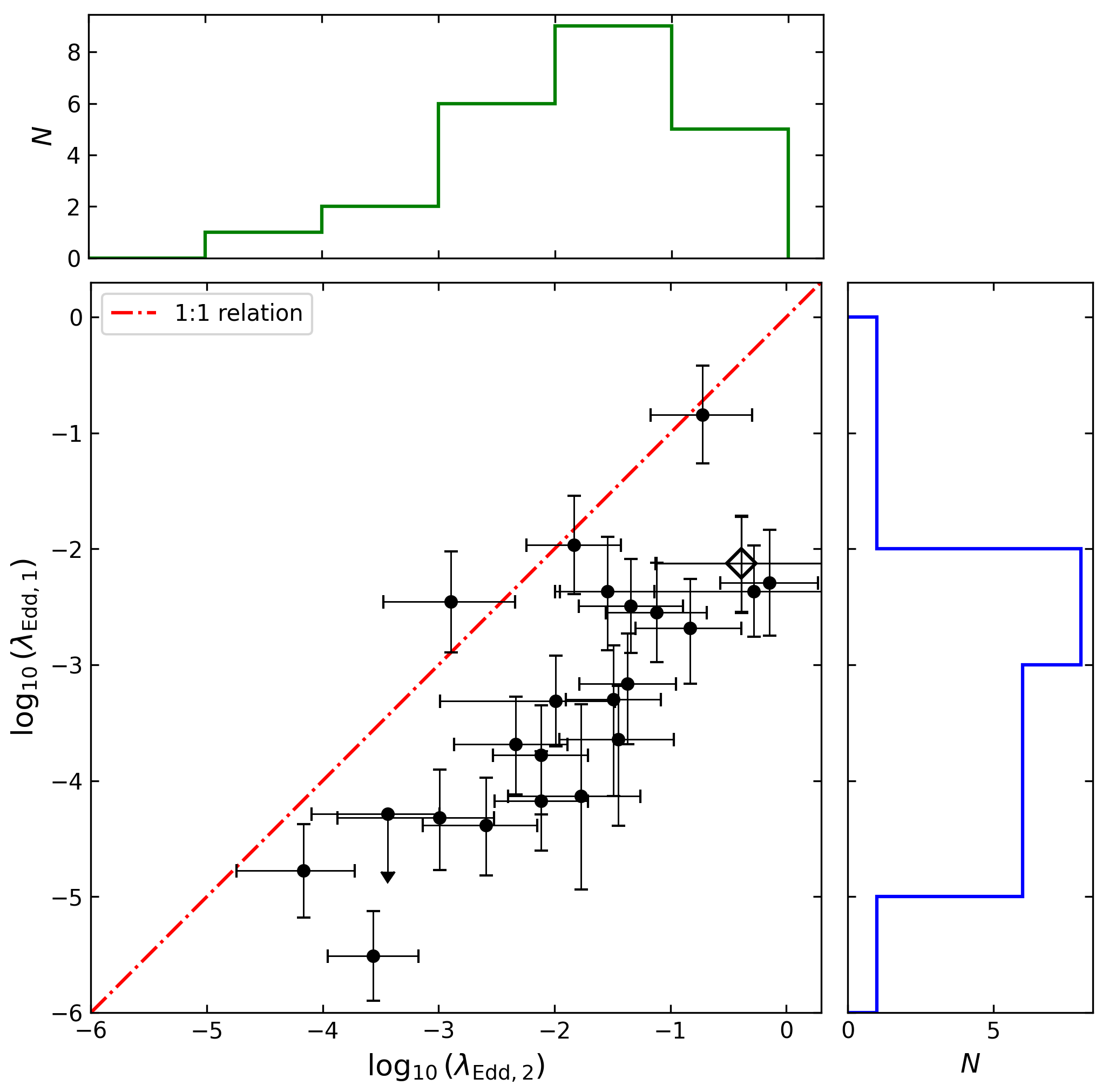} \caption{{\footnotesize Left-hand panel: mass of BH1 (blue squared data-points) and BH2 (black circled data-points) for AGN in the {AA} as a function of the projected spatial separation. Right-hand panel: Eddington ratio of the more massive BH of the pair versus the Eddington ratio of the less massive BH, computed through the spectral analysis. The red dashed line is the 1:1 relation. The blue and green histograms show the distribution (in terms of number of sources $N$) of $\log_{10}{(\lambda_{\rm Edd,1})}$ and $\log_{10}{(\lambda_{\rm Edd,2})}$, respectively. For both panels, the diamond data point is the IMBH J085125.8+393541.}} \label{fig6} \end{figure*}

Obscured AGN, when in pairs, tend to lie at the closest separations (see Section~\ref{sub.4.1} and Figure~\ref{fig3}). This suggests that, at close separations, galaxy nuclei may experience enhanced gas inflows, increasing the amount of material available to fuel SMBH accretion. In this scenario, AGN activity may be more efficiently triggered when the nuclei are closer together, leading to higher intrinsic luminosities. With the aim of investigating this behaviour, we searched whether dependencies exist between the AGN bolometric luminosity ($L_{\rm bol}$), BH mass ($M_{\rm BH}$), and Eddington ratio ($\lambda_{\rm Edd}$) as a function of $r_\mathrm{p}$. For each quantity $X$ (with $X = L_{\rm bol,1}$, $L_{\rm bol,2}$, $M_{\rm BH,1}$, $M_{\rm BH,2}$, $\lambda_{\rm Edd,1}$, or $\lambda_{\rm Edd,2}$, where 1 and 2 refer to the more massive, primary, and less massive, secondary, BH of each pair, respectively), if a correlation with projected separation was found using a Spearman test, we searched for a linear correlation using the following:

\begin{equation}
    \log_{10}{(X)}=\alpha\cdot \left(\frac{r_\mathrm{p}}{100\unit{kpc}}\right)+\beta\label{eq3}\,,
\end{equation}

\noindent where $\alpha$ and $\beta$ (and their associated errors) were estimated from the fit. Best-fit parameters are summarized in Table~\ref{tab0}. It is worth noticing that in our sample the source J085125.8+393541  is an intermediate-mass BH (IMBH) in early state of merger (diamond data-point in the left-hand panels of Figures \ref{fig5} and \ref{fig6}), with a BH mass $\log_{10}(M_{\rm BH}/{\rm M}_\odot)=4.6\pm0.6$. This source was already studied in previous works, where it was found to be hosted in a dwarf galaxy of stellar mass $\log_{10}(M_*/{\rm M}_\odot)=9.41$ and to have a BH mass of $\log_{10}(M_{\rm BH}/{\rm M}_\odot)=5.4$ \citep{reines2013} using the Full Width at Half Maximum (FWHM) of the H$\alpha$ line and its luminosity, following the approach from \citet{greene2005b} and \citet{bentz2013}. Moreover, this source resulted as an AGN also through a luminosity-based X-ray classification in \citet{wasleske}. Given the peculiar nature of this AGN, we considered it as an outlier point when performing our correlation tests.

When looking separately at the more massive BH of the pair (BH1) and the less massive one (BH2), a clear anti-correlation between the bolometric luminosity of BH2 and $r_{\rm p}$ is found (see the upper left-hand panel of Figure~\ref{fig5}), with a Spearman coefficient $\rho\sim-0.77$ and a $p$-value of $\sim2\times10^{-5}$. Conversely, the bolometric luminosity of BH1 ($L_{\rm bol,1}$) remains somewhat flat during the merger phases (see the upper right-hand panel of Figure~\ref{fig5}) with $\rho\sim-0.23$ and a $p$-value of $\sim0.29$ (no correlation is found)\footnote{Note that for $L_{\rm bol,2}$ we omit the results obtained by excluding the outlier point, as the correlation significance shows no appreciable changes.}. Given the significance of the correlation found for $L_{\rm bol,2}$ with $r_\mathrm{p}$, we then searched for a linear relation using Equation~\eqref{eq3}. The linear fit between $\log_{10}{(L_{\rm bol,2})}$ and $r_\mathrm{p}$ led us to:

\begin{equation}
    \log_{10}{\left(\frac{L_{\rm bol,2}}{\unit{erg\ s^{-1}}}\right)}=(-2.2\pm0.4)\cdot\left(\frac{r_\mathrm{p}}{100\unit{kpc}}\right)+(44.60\pm0.24)\,,\label{eq4}
\end{equation}

\noindent with a standard deviation at $\sigma_{\rm res}=0.56\unit{dex}$. This shows that in the late state of merger AGN could reach luminosities up to more than three orders of magnitude higher than that of AGN in the early state of merger. Moreover, there is no evidence for a larger  BH mass  with decreasing $r_{\rm p}$ both for BH1 and BH2 (Spearman coefficients $\rho\sim-0.01$ and $\rho\sim-0.11$, respectively), as shown in the left-hand panel of Figure~\ref{fig6}. This suggests that the increase in bolometric luminosity of BH2 is not driven by a systematically higher-mass BH population at the smallest separations, but instead may reflect enhanced accretion processes triggered by nuclei interaction, in agreement with the results of \citet{derosa}. This result shows that galaxy mergers in our sample play a role in triggering the AGN activity of the most luminous AGN. The best-fit for $L_{\rm bol,2}$ and its uncertainty region at $\pm3\sigma$ are shown in the upper left-hand panel of Figure~\ref{fig5} as a red dashed line and as a yellow shaded band, respectively.\\
\noindent{In order to exclude possible biases in relation~\eqref{eq4} due to selection effects at the closest separations, we also performed simulations to quantify the potential number of missing low-luminosity AGN in our sample. This is particularly relevant for the late stage of merger regime, in which a higher fraction of obscured AGN is expected, as described in Sections~\ref{sub.4.3} and \ref{sub.4.1}. This could lead to the lack of low-luminosity AGN that are not detected with XMM-Newton and {Chandra}, because their X-ray fluxes are below the sensitivity limit of the available observations (see Section~\ref{sub.2.1}). Our simulations (described in detail in Appendix~\ref{sec:appendixc}) revealed that the potential lack of low-luminosity AGN (with respect to all the detected AGN in the same bin of {$r_\mathrm{p}$}) corresponds to less than {$\sim$$8$\%} in the late and middle stage of merger, and to less than {$\sim$$5$\%} in the early stage of merger. This is clearly visible in Figure~\ref{figure_c1}, where the increase of bolometric luminosity in the first bin of {$r_\mathrm{p}$} after imposing flux cuts to match our observed sample is marginal and not significant (less than {$0.1\unit{dex}$}). This further check suggests that the anti-correlation between {$\log_{10}{(L_{\rm bol,2})}$} and {$r_\mathrm{p}$} is due to physical mechanisms occurring during the merger and not to a selection effect.}

In order to investigate the way through which two galaxies in a pair co-evolve, we examined the Eddington ratios of our dual AGN, a parameter which is crucial to study the accretion properties of the SMBH \citep{salpeter1964, lynden1969, shakura1973, soltan1982, rees1984} and its growth \citep{heckman2004, aird2012}. First, we compared the Eddington ratios between the more massive SMBH ($\lambda_{\rm Edd,1}$) and the less massive SMBH ($\lambda_{\rm Edd,2}$) within each pair. As was found also in previous studies (e.g., \citealt{bernhard2018, aggarwal2024}), the less massive BH of each pair shows, in the vast majority of cases, the highest $\lambda_{\rm Edd}$, which is indicative of a higher accretion occurring in the less massive SMBH (see Figure~\ref{fig6}). In fact, what arises when comparing sources 1 and 2 is the different peak of the respective Eddington ratio distributions (see the blue and green histograms in Figure~\ref{fig6}). The mean values for $<\log_{10}{(\lambda_{\rm Edd,1})}>$ and $<\log_{10}{(\lambda_{\rm Edd,2})}>$ are $\sim-3.1$ and $\sim-1.8$, with standard deviation $\sigma_{\rm std}=1.07$ and $\sigma_{\rm std}=1.05$, respectively.

To investigate in which way (if any) the mergers impact the accretion onto SMBHs in pairs, we then searched for a dependence of the Eddington ratios as a function of the projected spatial separation between the two nuclei of each pair. As a result, in Figure~\ref{fig5}, the Eddington ratios of sources 2 and 1 (lower left-hand and lower right-hand panel, respectively) are shown as a function of the projected spatial separation of the pair. Figure~\ref{fig5} shows an increase of $\lambda_{\rm Edd,2}$ as the separation between the two nuclei decreases. On the other hand, the Eddington ratios of sources 1 do not show this trend occurring during the merger phases. This is in agreement with the increase of bolometric luminosity of source 2 with decreasing $r_\mathrm{p}$, and with the flat $\log_{10}(L_{\rm bol,1})$--$r_\mathrm{p}$ relation, shown in the upper panels of Figure~\ref{fig5}. Considering that in our dual-AGN sample the majority of mergers are minor mergers (i.e., $M_{\rm BH,2}/M_{\rm BH,1}\leq0.25$), this result is in agreement with the hydrodynamical simulations by \citet{capelo2015}. In those simulations, a clear dependence of the bolometric luminosity and Eddington ratio on separation was observed, especially for minor mergers, albeit at different length-scales, with the decrease being confined at $\lesssim 10$-kpc scales. Not accounting for several other possible differences between observed and simulated galaxies (e.g., morphology, environment, orbital parameters), we note that the mass of the simulated BHs was in the range 0.4--$4 \times 10^6$~M$_{\odot}$, whereas the observed BHs mostly sit in the range $\sim$$10^7$--$10^9$~M$_{\odot}$ (see the left-hand panel of Figure~\ref{fig6}), approximately two orders of magnitude more massive. By presuming that also the galactic mass would increase by a similar factor \citep[e.g.,][]{Reines2015} and adopting a simple mass-radius quadratic relation \citep[e.g.,][]{Schulz2017}, we infer galactic sizes that would be larger by a factor of $\sim$10. Assuming that a galaxy is significantly perturbed only when the separation $R$ between the two galaxies (of masses $m$ and $M$, with $m < M$) is of the order of the tidal radius ($r_{\rm tidal} \simeq R (m/3M)^{1/3}$, for $m \ll M$), and assuming the same galaxy mass ratio of \citet{capelo2015}, we can surmise that simulations of these larger systems would show an interaction between the two galaxies starting \citep[and progressing; including the hydrodynamical torque effects caused by the expanded ram-pressure shock region, see][]{capelodotti2017} at larger distances, by a factor of $\sim$10, thus stretching the simulated luminosity-distance relation by the same factor and hence resembling the observed relation.

The Spearman test between $\lambda_{\rm Edd,1}$, $\lambda_{\rm Edd,2}$ and $r_\mathrm{p}$  is indicative of a moderate anti-correlation between $\lambda_{\rm Edd,2}$ and $r_\mathrm{p}$ (both including and excluding the outlier), which is not found for $\lambda_{\rm Edd,1}$ with $r_\mathrm{p}$ (see Table \ref{tab0}). By performing a fit on the $\lambda_{\rm Edd,2}$--$r_\mathrm{p}$ relation using Equation~\eqref{eq3}, if we include in the fit procedure the outlier IMBH J085125.8+393541, a linear anti-correlation is found with a Pearson coefficent of $\rho\sim-0.52$ and a standard deviation at $\sigma_{\rm res}=0.897\unit{dex}$, with a $p$-value of $\sim0.01$. If the outlier is excluded, we obtain instead $\rho\sim-0.67$, $\sigma_{\rm res}\sim0.765$, and a $p$-value of $\sim7\times10^{-4}$ (see Table \ref{tab0}). In the following, we only show the best-fit relation for $\log_{10}{(\lambda_{\rm Edd,2})}$ versus $r_\mathrm{p}$ obtained by excluding the outlier point (since it is consistent with the one obtained by including it and provides a higher statistical significance):

\begin{equation}
\log_{10}{(\lambda_{\rm Edd,2})}=(-2.3\pm0.6)\cdot\left(\frac{r_{\rm p}}{100\unit{kpc}}\right)+(-0.78\pm0.33)\,.
\end{equation}

\begin{table*}
    \centering
    \caption{{\footnotesize Results of the best fit performed to correlate $L_{\rm bol}$, $M_{\rm BH}$, and $\lambda_{\rm Edd}$ with $r_\mathrm{p}$ for BH2 and BH1 (using Equation~\ref{eq3}). For $\lambda_{\rm Edd,2}$, both the results including and excluding the outlier (out.) IMBH J085125.8+393541 are shown.}}\label{tab0}
    \footnotesize
    \tabcolsep=3pt
    \label{tab:results}
    \begin{tabular}{l cccccc}
    \hline
    \noalign{\smallskip}
    $X$ & $^{(\text{a})}\alpha$ & $^{(\text{b})}\beta$ & $^{(\text{c})}\sigma_{\rm res}$ & $^{(\text{d})}\rho$ & $^{(\text{e})}p$-value & $^{(\text{f})}$Mean \\
    & & & (dex) & & \\
    \noalign{\smallskip}
    \hline
    \noalign{\smallskip}
    $L_{\rm bol,2}$ & $-2.2\pm0.4$ & $44.60\pm0.24$ & $0.560$ & $-0.77$ & $2\times10^{-5}$ & $2\times10^{44}$~erg~s$^{-1}$\\
    $^*L_{\rm bol,1}$ & - & - & - & $-0.22$ & $0.11$ & $4\times10^{43}$~erg~s$^{-1}$\\
    $^*M_{\rm BH,2}$ & - & - & - & $-0.15$ & $0.49$ & $7.8\times10^7\unit{M_\odot}$\\
    $^*M_{\rm BH,1}$ & - & - & - & $-0.02$ & $0.94$ & $3.4\times10^8\unit{M_\odot}$ \\
    $\lambda_{\rm Edd,2}$ (out.) & $-1.8\pm0.6$ & $-0.92\pm0.38$ & $0.897$ & $-0.52$ & $0.01$ & -1.8\\
    $\lambda_{\rm Edd,2}$ (no out.) & $-2.3\pm0.6$ & $-0.78\pm0.33$ & $0.765$ & $-0.67$ & $7\times10^{-4}$ & -1.9\\
    $^*\lambda_{\rm Edd,1}$ & - & - & - & $-0.16$ & $0.46$ & -3.1\\
    \noalign{\smallskip}
    \hline
    \end{tabular}
    \vspace{1mm}
    \caption*{{\footnotesize (a): slope of the best-fit relation; (b): intercept of the best-fit relation; (c): standard deviation; (d): Spearman/Pearson coefficient. For the parameters for which a linear best-fit was found, $\rho$ corresponds to the Pearson coefficient, while it corresponds to the Spearman coefficient for parameters where no correlation was found with $r_\mathrm{p}$ (these parameters are marked with $^*$); (e): $p$-value. (f): mean value of the quantity.}}
\end{table*}

The best fit for $\lambda_{\rm Edd,2}$ and its uncertainty region at $\pm3\sigma$ are shown in the lower left-hand panel of Figure~\ref{fig5} as a red dashed line and as a yellow shaded band, respectively, if we consider the outlier. We also show the best fit and uncertainty region for $\lambda_{\rm Edd,2}$ as a black dotted line and grey shaded band in the lower left-hand panel of Figure~\ref{fig5}, if the outlier is excluded. This result suggests that the accretion processes for the less massive SMBH may be more powered the more it gets closer to its (more massive) companion. Such accretion processes may be triggered by the inflowing of gas onto the less massive SMBH of the pair caused by gravitational and hydrodynamical torques due to the presence of the more massive companion. This scenario is in agreement with the higher obscuration found in the late state of merger via the X-ray spectral analysis (see Section~\ref{sub.4.1}), if we assume that part of the gas which is obscuring the nucleus is also powering the SMBH accretion. These findings are in agreement with hydrodynamical simulations of galaxy mergers in which it was found that the AGN activity and obscuration increase during the merger stages \citep{capelo2015,capelo2017,blecha2018}.

\section{Summary and conclusions}\label{sec.5}

We investigated the properties of nearby ($z<0.1$) X-ray-selected dual AGN and AGN-galaxy pairs with XMM-Newton and Chandra ($r_{\rm p}=1$--$100\unit{kpc}$, $\Delta z<10^{-3}$). We performed spectral extraction and analysis in the X-ray band for 52 AGN in the AGN-AGN sample, and X-ray spectral analysis for 71 AGN in the AGN-galaxy sample, as well as optical spectroscopy analysis for 46 AGN in the AGN-AGN sample. From multi-wavelength spectral analysis, we derived key physical quantities that characterise AGN in pairs, such as the line-of-sigth obscuration, bolometric luminosity, and  Eddington ratio. We additionally performed a statistical comparison between the dual-AGN sample and a control sample of inactive galaxies in pairs and compared the properties of our dual-AGN sample with those of a sample of isolated AGN. 

\begin{itemize}

    \item Amongst all the AGN X-ray-selected (see Section~\ref{sec.2}), $\sim$10\% of them are found to be in a pair (with either a galaxy or an AGN as a companion). The fraction of AGN paired with a galaxy is $\sim$6\%, while the fraction of dual AGN is $\sim$4\% (see Figure~\ref{fig2}). The latter fraction reduces to $\sim$2\% when looking at AGN in the late state of merger ($r_{\rm p}<30\unit{kpc}$).
    
    \item Using a threshold on XMM-Newton hardness ratios in four energy bands (i.e., HR3$\,>0$ and HR4$\,>0.2$), we found that $\gtrsim  44\%$ of isolated AGN are absorbed with $N_{\rm H}>10^{22}\unit{cm^{-2}}$. It is worth noticing  that this fraction should be regarded as a lower limit, since our selection criteria do not allow us to identify the whole population of Compton-thick sources (amongst those showing HR3$\,<0$ and HR4$\,<0.2$).
    
    \item We found that $\sim$55\% of AGN in pairs are absorbed with $N_{\rm H}>10^{22}\unit{cm^{-2}}$ with a mean value of $\sim$$1.8\times10^{23}\unit{cm^{-2}}$.  This fraction is $\sim$60\% for AGN-AGN pairs and $\sim$50\% for AGN-galaxy pairs. The fraction of Compton-thick AGN in pairs is $\sim$6\%.  Overall, our results suggest that the fraction of obscured AGN in pairs is higher with respect to isolated AGN, although the comparison remains uncertain, because we only have a lower limit for the fraction of absorbed isolated AGN, which could therefore be higher than $44\%$.
    
    \item We found that the fraction of absorbed AGN in pairs ($N_{\rm H}>10^{22}\unit{cm^{-2}}$) increases as galaxy interactions progress. Specifically, the fractions are $\sim$65\% ($\sim$70\% for AGN-AGN pairs and $\sim$61\% for AGN-galaxy pairs) in the late state of merger, $\sim$56\% ($\sim$70\% for AGN-AGN pairs and $\sim$45\% for AGN-galaxy pairs) in the middle state of merger, and $\sim$35\% ($\sim$30\% for AGN-AGN pairs and $\sim$39\% for AGN-galaxy pairs) in the early state of merger (see Figure~\ref{fig3}). Although we did not find a clear correlation between $N_\mathrm{H}$ and the projected spatial separation, these results indicate that obscuration becomes more common as galaxy pairs evolve toward late merger stages.  
    
    \item We quantified the excess of AGN pairs with respect to inactive galaxy pairs in bins of projected spatial separation. In each bin, the excess is defined as $f^{\rm AGN}_{\rm bin}/f^{\rm gal}_{\rm bin}$, where $f^{\rm AGN}_{\rm bin}$ and $f^{\rm gal}_{\rm bin}$ are the fractions of AGN and galaxy pairs in each bin normalized by the total number of AGN or galaxies in pairs plus isolated systems (i.e., number of AGN/galaxies in the bin over the total number of AGN/galaxies in pairs+isolated). {We estimated the excess for: all AGN in pairs; only obscured AGN in pairs; and only un-obscured AGN in pairs}. Our analysis revealed that AGN pairs are{, on average,} more common than inactive galaxy pairs, {but a significant correlation between AGN excess and {$r_\mathrm{p}$} was only found for obscured AGN (see Figure~\ref{fig4}). This suggests that mergers may have an impact on AGN triggering, especially for obscured AGN, and reflects the higher obscuration regime for AGN in close pairs.}
    
    \item We investigated the relation between the bolometric luminosity of AGN in AGN-AGN pairs with respect to the projected spatial separation. We found a clear increase of the  bolometric luminosity of the less massive BH of each pair as the interaction progresses. Specifically, we found a significant anti-correlation between $L_{\rm bol,2}$ and $r_{\rm p}$ (with a Pearson coefficient of $r\sim-0.77$ and a $p$-value $\sim2\times10^{-5}$, see the upper left-hand panel of Figure~\ref{fig5} and Table~\ref{tab0}). {We also ensured through simulations that this anti-correlation is not driven by a possible bias in AGN selection due to their higher obscuration at closest separations (with obscured fractions found through our analysis, see Figure \ref{fig3}), but it could be due to physical mechanisms occurring during the merger}. This anti-correlation is not found for the more massive BHs of the pairs (see the upper right-hand panel of Figure~\ref{fig5}). These results suggest that mergers preferentially enhance the accretion onto the less massive black hole, leading to more luminous secondary AGN at the closest separations. The flat trend between the BH mass and projected spatial separation (see the left-hand panel of Figure~\ref{fig6}) suggests that the increase of bolometric luminosity for BH2 is possibly due to enhanced accretion processes triggered by nuclei interaction.
    
    \item We found that, in the majority of AGN pairs, the less massive BH accretes at higher Eddington ratio $\lambda_{\rm Edd}$ than its companion (see Figure~\ref{fig6}). There is a clear evidence of an increase of the Eddington ratio for the less massive BH of the pair as the projected spatial separation between the nuclei decreases (see the lower left-hand panel of Figure~\ref{fig5}), consistent with the observed rise of $L_{\rm bol,2}$ and the lack of any trend between $M_{\rm BH,2}$ with $r_{\rm p}$. Conversely,  the Eddington ratio of the more massive BH of the pair does not show an increase during the merger phases (see the lower right-hand panel of Figure~\ref{fig5}). Taken together, our findings suggest  that, in minor mergers, galaxy interactions may have an important role in triggering and enhancing the AGN activity of the less massive BH in the pair, leading to stronger accretion and higher luminosity as the nuclei approach each other.
    \end{itemize}

\noindent{Data availability}\\
The high-level data underlying this article are extracted through standard processing from raw data stored in public archives (SDSS, XMM-Newton, and Chandra), and will be shared on reasonable request to the corresponding author.
    
\begin{acknowledgements}
We would like to thank the referee for the time and for the constructive comments, which have helped us to significantly improve the quality of the manuscript. We acknowledge financial contribution from INAF Large Grant ``The Quest for dual and binary massive black holes in the gravitational wave era'' (Bando Ricerca Fondamentale INAF 2024) and  from INAF Large Grant ``Dual and binary supermassive black holes in the multi-messenger era: from galaxy mergers to gravitational waves'' (Bando Ricerca Fondamentale INAF 2022). We acknowledge financial support from the ASI-INAF agreement n. 2024-36-HH.1-2025 and  n. 2025-29-HH.0. PRC acknowledges support from the Swiss National Science Foundation under the Sinergia Grant CRSII5\_213497 (GW-Learn) and fruitful discussions with Massimo Dotti. EB acknowledges the support of  the INAF GO grant ``A JWST/MIRI MIRACLE: Mid-IR Activity of Circumnuclear Line Emission'' and of the ``Ricerca Fondamentale 2024'' INAF program (mini-grant 1.05.24.07.01). We acknowledge support from PRIN-MUR project ``PROMETEUS''  financed by the European Union -  Next Generation EU, Mission 4 Component 1 CUP B53D23004750006 and C53D2300080006. This research has made use of data obtained from the 4XMM XMM-Newton serendipitous source catalogue compiled by the 10 institutes of the XMM-Newton Survey Science Centre selected by ESA. This research has made use of data obtained from the Chandra Data Archive, observations made by the Chandra X-ray Observatory and published previously in cited articles. This publication has made use of data products from the Wide-field Infrared Survey Explorer, which is a joint project of the University of California, Los Angeles, and the Jet Propulsion Laboratory/California Institute of Technology, funded by the National Aeronautics and Space Administration. Funding for SDSS-III has been provided by the Alfred P. Sloan Foundation, the Participating Institutions, the National Science Foundation, and the U.S. Department of Energy Office of Science. The SDSS-III web site is http://www.sdss3.org/. SDSS-III is managed by the Astrophysical Research Consortium for the Participating Institutions of the SDSS-III Collaboration including the University of Arizona, the Brazilian Participation Group, Brookhaven National Laboratory, Carnegie Mellon University, University of Florida, the French Participation Group, the German Participation Group, Harvard University, the Instituto de Astrofisica de Canarias, the Michigan State/Notre Dame/JINA Participation Group, Johns Hopkins University, Lawrence Berkeley National Laboratory, Max Planck Institute for Astrophysics, Max Planck Institute for Extraterrestrial Physics, New Mexico State University, New York University, Ohio State University, Pennsylvania State University, University of Portsmouth, Princeton University, the Spanish Participation Group, University of Tokyo, University of Utah, Vanderbilt University, University of Virginia, University of Washington, and Yale University. FR acknowledges support from ELSA. ELSA Euclid Legacy Science Advanced analysis tools” (Grant Agreement no. 101135203) is funded by the European Union. Views and opinions expressed are, however, those of the author(s) only and do not necessarily reflect those of the European Union or Innovate UK. Neither the European Union nor the granting authority can be held responsible for them. UK participation is funded through the UK Horizon guarantee scheme under Innovate UK grant 10093177.
\end{acknowledgements}

  \bibliographystyle{aa}
\bibliography{bibliography}

\begin{appendix}

\section{Spectral analysis: tables of results}\label{sec:appendix}

Tables~\ref{tab1} and \ref{tab3} report the results of X-ray and optical spectral analysis on single sources as described in Section~\ref{sec.3}.
\begin{table*}[ht]
    \centering
    \caption{Results of the X-ray spectral analysis described in Section~\ref{sec.3}.}\label{tab1}
    \resizebox{1\textwidth }{!}{
    \begin{tabular}{cccccccccc}
    \hline
    \multicolumn{10}{c}{}\\
       $^{\text{(a)}}$IAUNAME  &  $^{\text{(b)}}z$ & $^{\text{(c)}}r_{\rm p}$ & $^{\text{(d)}}\Gamma^H$ & $^{\text{(e)}}\Gamma^S$ & $^{\text{(f)}}N_{\rm H}$ & $^{\text{(g)}}F^{\rm Obs}_{\rm X}$ & $^{\text{(h)}}L^{\rm Unabs}(2$--$10\unit{keV})$ & $^{\text{(i)}}$Net counts & $^{\text{(l)}}$Opt. spec. \\
       & & ($\unit{kpc}$) & & & ($10^{22}\unit{cm^{-2}}$) & ($10^{-14}\unit{erg\ s^{-1}\ cm^{-2}}$) & ($10^{41}\unit{erg\ s^{-1}}$) & (cts) &  \\
    \multicolumn{10}{c}{}\\
       \hline
    \multicolumn{10}{c}{}\\
    2CXO J135602.8+182218 & 0.051 & 3.9 & $1.8\pm0.2$ & * & $54^{+16}_{-13}$ & $32\pm4$ & $103^{+17}_{-14}$ & {$1267\pm36$} & \ding{55}\\
    2CXO J135602.6+182217 & 0.051 & 3.9 & $1.9^f$ & $2.5^{+0.5}_{-0.4}$ & $34^{+8}_{-6}$ & $22^{+3}_{-10}$ & $48\pm6$ & {$324\pm18$} & \ding{55}\\
    { 2CXO J121346.0+024841}$^\ddagger$ & 0.073 & 4.8 & $2.1\pm0.6$ & - & - & $0.9\pm0.2$ & $1.3\pm3$ & {$15\pm5$} & \ding{51}\\
    { 2CXO J121345.9+024838}$^\ddagger$ & 0.073 & 4.8 & $1.9^f$ & - & $<0.6$ & $0.7\pm0.2$ & $1.1\pm4$ & {$11\pm4$} & \ding{55}\\
    { 2CXO J133817.8+481640}$^\ddagger$ & 0.028 & 6.4 & $1.9^f$ & - & $6.8\pm3.8$ & $24.6\pm2$ & $4.0$ & {$205\pm15$} & \ding{51}\\
    { 2CXO J133817.3+481632}$^\ddagger$ & 0.028 & 6.4 & $1.6\pm0.3$ & - & $>200$ & $2.2\pm0.9$ & $130^{+270}_{-80}$ & {$95\pm11$} & \ding{51}\\
    2CXO J121417.8+293143 & 0.064 & 7.9 & $1.9^f$ & $2.0^{+1.2}_{-1.3}$ & $25^{+13}_{-7}$ & $140\pm20$ & $370\pm60$ & {$215\pm15$} & \ding{51}\\
    { 2CXO J090714.6+520350}$^\ddagger$ & 0.060 & 8.9 & $1.5\pm0.7$ & - & $2.5^{+1.4}_{1.1}$ & $13\pm1$ & $16.0\pm0.1$ & {$121\pm12$} & \ding{51}\\
    { 2CXO J090714.4+520343}$^\ddagger$ & 0.060 & 8.9 & $1.9^f$ & - & $18^{+16}_{-12}$ & $6.4\pm1.1$ & $16\pm3$ & {$42\pm7$} & \ding{51}\\
    2CXO J133440.7-232645 & 0.033 & 10.5 & $1.1^{+1.2}_{-0.4}$ & - & $<1.5$ & $40\pm7$ & $9.8^{+2.2}_{-1.8}$ & {$91\pm10$} & \ding{55}\\
    2CXO J133439.6-232647 & 0.034 & 10.5 & $1.9^f$ & * & $15^{+3}_{-2}$ & $300\pm30$ & $170^{+20}_{-10}$ & {$366\pm20$} & \ding{55}\\
    2CXO J032512.9+404152 & 0.048 & 12.0 & $2.0^f$ & - & $0.5\pm0.3$ & $16^{+3}_{-2}$ & $9.5^{+0.6}_{-1.3}$ & {$152\pm12$} & \ding{55}\\
    2CXO J032512.2+404202 & 0.048 & 12.0 & $1.1\pm0.4$ & * & $6.7^{+1.4}_{-1.0}$ & $390^{+20}_{-30}$ & $280^{+10}_{-20}$ & {$957\pm31$} & \ding{55}\\
    2CXO J110018.0+100257 & 0.036 & 12.6 & $2.4^{+0.6}_{-0.5}$ & - & $<0.1$ & $3.0\pm0.5$ & $0.9\pm0.2$ & {$92\pm10$} & \ding{51}\\
    2CXO J145051.5+050652 & 0.028 & 12.6 & $1.9^f$ & $3.3\pm0.3$ & $182^{+167}_{-92}$ & $24\pm10$ & $240\pm100$ & {$224\pm15$} & \ding{51}\\
    2CXO J012357.0-350409 & 0.019 & 13.9 & $1.4\pm0.5$ & - & $<0.1$ & $6.2\pm1.4$ & $0.5\pm0.1$ & {$62\pm8$} & \ding{55}\\
    2CXO J012354.3-350355 & 0.019 & 13.9 & $1.5^f$ & - & $0.9\pm0.1$ & $760^{+40}_{-20}$ & $67\pm2$ & {$4964\pm70$} & \ding{55}\\
    2CXO J020920.8-100759 & 0.013 & 15.0 & $1.9^f$ & * & $43^{+24}_{-13}$ & $32\pm7$ & $5.7\pm1.3$ & {$81\pm9$} & \ding{51}\\
    2CXO J020924.5-100809 & 0.013 & 15.0 & $2.2^f$ & $3.1^{+1.6}_{-1.2}$ & $49^{+27}_{-17}$ & $31\pm7$ & $7.3\pm1.5$ & {$141\pm12$} & \ding{55}\\
    2CXO J181611.6+423937 & 0.041 & 23.2 & $1.9^f$ & * & $3.6^{+0.4}_{-0.2}$ & $240\pm10$ & $130\pm10$ & {$4308\pm66$} & \ding{55}\\
    2CXO J181609.3+423922 & 0.042 & 23.2 & $1.9^f$ & - & $46^{+14}_{-10}$ & $33\pm6$ & $61\pm10$ & {$137\pm12$} & \ding{55}\\
    2CXO J132809.9-271954 & 0.042 & 28.0 & $1.8^{+0.3}_{-0.2}$ & - & $0.8\pm0.2$ & $280\pm20$ & $120\pm10$ & {$770\pm28$} & \ding{55}\\
    2CXO J132807.7-272006 & 0.042 & 28.0 & $1.3^f$ & * & $3.9^{+3}_{-2}$ & $200\pm30$ & $100\pm10$ & {$412\pm20$} & \ding{55}\\
    4XMM J094548.3-142204 & 0.008 & 29.0 & $1.6\pm0.1$ & - & $<0.03$ & $28\pm1$ & $0.4\pm0.2$ & $2519\pm51$ & \ding{55}\\
    2CXO J094541.9-141934 & 0.008 & 29.0 & $1.1\pm0.1$ & - & $0.2\pm0.1$ & $310\pm10$ & $4.4\pm0.2$ & {$4747\pm70$} & \ding{55}\\
    { 4XMM J094554.4+423818}$^\ddagger$ & 0.075 & 30.3 & $1.9^f$ & - & $24^{+8}_{-6}$ & $19\pm2$ & $69\pm3$ & $1275\pm37$ & \ding{51}\\
    { 4XMM J094554.4+423840}$^\ddagger$ & 0.075 & 30.3 & $2.4\pm0.1$ & - & $<0.02$ & $50\pm3$ & $70\pm3$ & $21450\pm150$ & \ding{51}\\
    { 4XMM J120445.3+311130}$^{\ddagger}$ & 0.025 & 31.0 & $1.9^f$ & - & $24^{+150}_{-12}$ & $3.0\pm1.1$ & $1.1^{+0.7}_{-0.6}$ & $154\pm20$ & \ding{51}\\
    { 4XMM J120443.3+311037}$^\ddagger$ & 0.025 & 31.0 & $1.7\pm0.4$ & - & $3.7\pm0.5$ & $263\pm6$ & $31\pm1$ & $16517\pm137$ & \ding{51}\\
    4XMM J121958.9-355735 & 0.057 & 35.6 & $3.3^{1.4}_{-1.2}$ & * & $>30$ & $3.0\pm1.5$ & $16\pm8$ & $211\pm17$ & \ding{55}\\
    4XMM J121957.3-355801 & 0.058 & 35.6 & $1.8^{+0.6}_{-0.5}$ & - & $<0.01$ & $1.8\pm0.4$ & $1.4\pm0.4$ & $181\pm16$ & \ding{55}\\
    2CXO J091338.8-101919 & 0.055 & 38.8 & $3.1^{+0.8}_{-0.7}$ & - & $<0.01$ & $0.8\pm0.2$ & $0.6\pm0.1$ & {$82\pm9$} & \ding{55}\\
    { 4XMM J103855.9+392157}$^\ddagger$ & 0.055 & 42.9 & $1.9^f$ & - & $>100$ & $1.6\pm1.0$ & $70$ & $182\pm14$ & \ding{51}\\
    { 4XMM J103853.3+392151}$^\ddagger$ & 0.055 & 42.9 & $1.6\pm0.1$ & - & $<0.03$ & $55\pm5$ & $42\pm2$ & $6534\pm80$ & \ding{51}\\
    { 4XMM J100135.8+033648}$^\ddagger$ & 0.043 & 45.3 & $1.9^f$ & - & $16^{+31}_{-10}$ & $6.3\pm1.1$ & $4.6\pm2$ & $260\pm29$ & \ding{51}\\
    { 4XMM J162644.4+142253}$^\ddagger$ & 0.048 & 51.3 & $1.9^f$ & - & $6^{+12}_{-4}$ & $1.6\pm0.8$ & $1.2\pm0.8$ & $162\pm15$ & \ding{51}\\
    { 4XMM J162640.9+142243}$^\ddagger$ & 0.048 & 51.3 & $1.9^f$ & - & $67^{+150}_{-50}$ & $2.3\pm1$ & $8^{+3}_{-7}$ & $364\pm20$ & \ding{51}\\
    { 4XMM J145627.4+211956}$^\ddagger$ & 0.044 & 59.2 & $1.9\pm0.9$ & - & $75^{+28}_{-23}$ & $8.6\pm2.0$ & $37^{+2}_{-13}$ & $602\pm25$ & \ding{51}\\
    { 4XMM J145631.3+212030}$^\ddagger$ & 0.044 & 59.2 & $1.9^f$ & - & $>100$ & $2.3\pm0.5$ & $70$ & $459\pm32$ & \ding{51}\\
    4XMM J093402.8+100631 & 0.010 & 60.1 & $1.9^f$ & $4.2^{+2.6}_{-1.5}$ & $18^{+5}_{-4}$ & $25\pm4$ & $1.4\pm0.2$ & $455\pm25$ & \ding{51}\\
    4XMM J093346.0+100909 & 0.011 & 60.1 & $1.8^{+0.6}_{-0.3}$ & - & $<0.2$ & $5.4\pm0.7$ & $0.12\pm0.02$ & $838\pm32$ & \ding{51}\\
    2CXO J201010.8-481713 & 0.010 & 76.6 & $1.9^{+0.2}_{-0.1}$ & - & $<0.01$ & $15\pm2$ & $0.33\pm0.03$ & {$496\pm22$} & \ding{55}\\
    2CXO J200954.0-482246 & 0.010 & 76.6 & $1.6^f$ & - & $<0.08$ & $11\pm1$ & $0.24\pm0.02$ & {$1556\pm39$} & \ding{55}\\
    4XMM J165105.6-012748 & 0.041 & 82.3 & $1.8^f$ & * & $72^{+32}_{-23}$ & $31\pm5$ & $93^{+17}_{-15}$ & $765\pm34$ & \ding{55}\\
    4XMM J165105.8-012926 & 0.041 & 82.3 & $1.7^{+1.1}_{-0.3}$ & $1.4^{+1.9}_{-1.2}$ & $4.8^{+3.0}_{-1.3}$ & $78\pm5$ & $41\pm3$ & $1768\pm47$ & \ding{55}\\
    4XMM J232903.9+033159 & 0.017 & 87.8 & $2.3^f$ & * & $45^{+19}_{-14}$ & $28\pm5$ & $11\pm2$ & $1493\pm47$ & \ding{55}\\
    4XMM J232846.6+033041 & 0.016 & 87.8 & $2.0^f$ & - & $<0.01$ & $54\pm2$ & $3.1\pm0.1$ & $6343\pm90$ & \ding{55}\\
    { 4XMM J100236.6+324224}$^\ddagger$ & 0.051 & 88.1 & $2.2^{+0.9}_{-0.6}$ & - & $<0.4$ & $10\pm1$ & $4.7\pm1$ & $2359\pm93$ & \ding{51}\\
    { 4XMM J082321.6+042221}$^\ddagger$ & 0.031 & 88.7 & $3.0^{+0.5}_{-0.4}$ & - & $0.6^{+1.4}_{-0.3}$ & $10\pm1$ & $2.2^{+0.4}_{-0.8}$ & $8871\pm125$ & \ding{51}\\
    4XMM J102141.8+130551 & 0.076 & 94.4 & $1.9^f$ & - & $<0.05$ & $0.4\pm0.2$ & $0.6\pm0.3$ & $45\pm11$ & \ding{51}\\
    4XMM J102142.6+130654 & 0.077 & 94.4 & $1.7^{+0.6}_{-0.5}$ & - & $<0.1$ & $1.4\pm0.4$ & $2.0\pm0.5$ & $179\pm17$ & \ding{51}\\
    2CXO J085125.8+393541 & 0.041 & 94.6 & $1.6\pm0.5$ & - & $<0.01$ & $4.6\pm1$ & $1.8\pm0.4$ & {$61\pm8$} & \ding{51}\\
    \multicolumn{10}{c}{}\\
    \hline
    \end{tabular}}
    \vspace{1mm}
    \caption*{(a) Name of the source; (b) redshift; (c) projected spatial separation between the nuclei; (d) photon index of the primary component; (e) photon index of the soft power law; (f) hydrogen column density; (g) observed flux in the $2$--$10\unit{keV}$ band; (h) de-absorbed luminosity in the $2$--$10\unit{keV}$ band; (i) net counts of the spectrum in the $0.2$--$12\unit{keV}$ and {$0.5$}{--}{$8\unit{keV}$} bands for XMM-Newton and Chandra, respectively; (l) check for availability of SDSS optical spectrum. Sources marked with \ding{51}/\ding{55} have/do not have an optical spectrum available from SDSS. $\ddagger$ sources are the AGN in the \citet{derosa} sample. The superscripts $f$ for the values of $\Gamma^H$ and $\Gamma^S$ mean that the value was fixed either at a typical value of $1.9$ or at the best-fit value. When the soft photon index was fixed to the hard photon index value, it is marked with *. If the soft component was not included in the model, or it refers to a source from \citet{derosa}, we marked it with -.}
\end{table*}

\begin{table*}[ht]
    \centering
    \caption{Results of the optical spectral analysis described in Section~\ref{sec.3}.}\label{tab3}
     \resizebox{1\textwidth }{!}{
    \begin{tabular}{cccccccc}
    \hline
    \multicolumn{1}{c}{}\\
    $^{\text{(a)}}$IAUNAME & $^{\text{b}}F({\rm H}\alpha)$ & $^{\text{(c)}}$$F({\rm H}\beta)$ & $^{\text{(d)}}$$F([{\rm O}III])$ & $^{\text{(e)}}$$\log_{10}{(M_{\rm BH}/{\rm M}_\odot)}$ & $^{\text{(f)}}$$L_{\rm bol}$ & $^{\text{(g)}}$X-ray spec.\\
    & ($10^{-16}\unit{erg\ s^{-1}\ cm^{-2}})$ & ($10^{-16}\unit{erg\ s^{-1}\ cm^{-2}})$ & ($10^{-16}\unit{erg\ s^{-1}\ cm^{-2}}$) & & ($10^{42}\unit{erg\ s^{-1}}$) & \\
    \multicolumn{1}{c}{}\\
    \hline
    \multicolumn{7}{c}{}\\
    J133817.8+481640  & $(1.0\pm0.4)\times10^{3}$ & $386\pm4$ & $619\pm6$ & $8.9\pm0.8$ & $6.8\pm2.4$ & \ding{51}\\
    J133817.3+481632 & $467\pm5$ & $83\pm5$ & $537\pm7$ & $8.2\pm0.5$ & $350\pm220$ & \ding{51}\\
    J121418.2+293146 & $60\pm1$ & $22\pm1$ & $208\pm1$ & $8.0\pm0.4$ & $59\pm2$ & \ding{55}\\
    J121417.8+293143 & $511\pm5$ & $182\pm5$ & $2\times10^3\pm7$ & $7.5\pm1.7$ & $(1.9\pm0.3)\times10^3$ & \ding{51}\\
    J090714.4+520343 & $37\pm1$ & $10\pm1$ & $66\pm1$ & $7.9\pm0.4$ & $32\pm6$ & \ding{51}\\
    J090714.6+520350 & $59\pm1$ & $15\pm1$ & $49\pm1$ & $7.0\pm0.5$ & $55\pm2$ & \ding{51}\\
    J110018.0+100257 & $123\pm1$ & $15\pm1$ & $47\pm1$ & $8.1\pm0.4$ & $1.0\pm0.2$ & \ding{51}\\
    J110019.1+100250 & $39\pm1$ & $5.6\pm0.3$ & $40\pm1$ & $8.0\pm0.4$ & $94\pm15$ & \ding{55}\\
    J145050.6+050710 & $17\pm1$ & $2.5\pm0.3$ & $4.2\pm0.4$ & $8.0\pm0.4$ & $24\pm11$ & \ding{55}\\
    J145051.5+050652 & $442\pm2$ & $124\pm2$ & $821\pm2$ & $7.8\pm0.4$ & $(1.1\pm0.4)\times10^3$ & \ding{51}\\
    J120157.8+295945 & $6.0\pm0.4$ & $1.8\pm0.4$ & $5.9\pm0.5$ & $8.5\pm0.4$ & $9.5\pm7.3$ & \ding{55}\\
    J120157.8+295927 & $11\pm1$ & $2.9\pm0.3$ & $6.2\pm0.4$ & $6.7\pm0.5$ & $25\pm9$ & \ding{55}\\
    J094554.4+423840 & $348\pm2$ & $134\pm2$ & $74\pm2$ & $8.8\pm0.4$ & $212\pm9$ & \ding{51}\\
    J094554.4+423818 & $141\pm1$ & $38\pm1$ & $213\pm1$ & $7.3\pm0.4$ & $208\pm9$ & \ding{51}\\
    J120445.3+311130 & $17\pm1$ & $5.5\pm0.6$ & $17\pm1$ & $7.1\pm0.4$ & $1.2\pm0.6$ & \ding{51}\\
    J120443.3+311037 & $412\pm3$ & $124\pm3$ & $(1.4\pm0.4)\times10^3$ & $7.1\pm0.4$ & $74\pm1$ & \ding{51}\\
    J011834.4-000939 & $18\pm1$ & $4.3\pm0.5$ & $5.4\pm0.6$ & $8.1\pm0.4$ & $63\pm27$ & \ding{55}\\
    J011832.8-000936 & $2.8\pm0.3$ & $8.6\pm0.4$ & $68\pm1$ & $7.0\pm0.4$ & $35\pm1$ & \ding{55}\\
    J103853.3+392151 & $48\pm1$ & $18\pm1$ & $55\pm1$ & $8.2\pm0.5$ & $109\pm5$ & \ding{51}\\
    J103855.9+392157 & $30\pm1$ & $9.1\pm0.3$ & $138\pm1$ & $6.4\pm0.4$ & $212\pm1$ & \ding{51}\\
    J100133.4+033731 & $2.6\pm0.4$ & $1.1\pm0.4$ & $4.6\pm0.5$ & $7.7\pm0.4$ & $1.4\pm0.5$ & \ding{55}\\
    J100135.8+033648 & $71\pm1$ & $20\pm1$ & $80\pm1$ & $7.1\pm0.4$ & $6.7\pm3.6$ & \ding{51}\\
    J162644.4+142253 & $4.6\pm0.5$ & $1.1\pm0.5$ & $4.2\pm0.4$ & $7.3\pm0.4$ & $1.3\pm0.1$ & \ding{51}\\
    J162640.9+142243 & $43\pm1$ & $12\pm1$ & $181\pm1$ & $7.0\pm0.4$ & $13\pm12$ & \ding{51}\\
    J145631.3+212030 & $47\pm1$ & $9.9\pm0.8$ & $181\pm1$ & $7.1\pm0.4$ & $212\pm1$ & \ding{51}\\
    J145627.4+211956 & $13\pm1$ & $4.5\pm0.4$ & $41\pm1$ & $6.6\pm0.4$ & $93\pm33$ & \ding{51}\\
    J015235.9-083139 & $9.5\pm0.6$ & $2.7\pm0.6$ & $6.5\pm0.7$ & $8.2\pm0.4$ & $9.6\pm7.9$ & \ding{55}\\
    J015235.2-083233 & $29\pm1$ & $10\pm1$ & $39\pm1$ & $6.7\pm0.4$ & $18\pm1$ & \ding{55}\\
    J093346.0+100909 & $155\pm2$ & $44\pm2$ & $73\pm2$ & $8.4\pm0.4$ & $0.9\pm0.1$ & \ding{51}\\
    J093402.8+100631 & $17\pm1$ & $6.8\pm1.1$ & $25\pm1$ & $7.7\pm0.4$ & $1.6\pm0.2$ & \ding{51}\\
    J161211.2+293428 & $6.7\pm0.3$ & $1.9\pm0.3$ & $6.7\pm0.4$ & $8.6\pm0.4$ & $8.2\pm4.4$ & \ding{55}\\
    J161216.7+293423 & $28\pm1$ & $8.1\pm0.6$ & $63\pm1$ & $7.5\pm0.4$ & $32\pm9$ & \ding{55}\\
    J160501.3+174632 & $4.1\pm0.7$ & $0.9\pm0.7$ & $4.5\pm0.8$ & $8.6\pm0.4$ & $<2.8$ & \ding{55}\\
    J160508.1+174528 & $9.3\pm0.5$ & $2.6\pm0.5$ & $5.8\pm0.6$ & $8.0\pm0.4$ & $4.2\pm2.9$ & \ding{55}\\
    J134130.5-002512 & $33\pm1$ & $7.3\pm0.4$ & $5.0\pm0.5$ & $7.9\pm0.4$ & $109\pm21$ & \ding{55}\\
    J134133.4-002431 & $40\pm2$ & $13\pm1$ & $123\pm2$ & $7.5\pm0.4$ & $58\pm8$ & \ding{55}\\
    J100236.6+324224 & $11\pm2$ & $6.8\pm1.5$ & $3.1\pm1.3$ & $9.5\pm0.4$ & $6.8\pm1.5$ & \ding{51}\\
    J100230.8+324248 & $12\pm1$ & $3.9\pm0.7$ & $3.2\pm0.9$ & $9.0\pm0.4$ & $9.1\pm5.7$ & \ding{55}\\
    J082321.6+042221 & $9.2\pm0.8$ & $5.0\pm0.8$ & $4.1\pm1$ & $8.7\pm0.4$ & $2.7\pm1$ & \ding{51}\\
    J082329.9+042332 & $11\pm1$ & $2.7\pm0.4$ & $22\pm1$ & $7.2\pm0.4$ & $4.9\pm2.9$ & \ding{55}\\
    J102142.6+130654 & $52\pm2$ & $9.5\pm0.2$ & $4.4\pm0.3$ & $6.7\pm0.4$ & $2.4\pm0.6$ & \ding{51}\\
    J102141.8+130551 & $23\pm1$ & $4.8\pm0.2$ & $17\pm1$ & $6.6\pm0.5$ & $0.6\pm0.3$ & \ding{51}\\
    J085135.7+393523 & $24\pm1$ & $5.9\pm0.5$ & $18\pm1$ & $7.4\pm0.4$ & $22\pm7$ & \ding{55}\\ J085125.8+393541 & $38\pm1$ & $15\pm1$ & $50\pm3$ & $4.6\pm0.7$ & $2.1\pm0.5$ & \ding{51}\\
    J163102.7+394733 & $6.0\pm0.9$ & $2.2\pm0.9$ & $3.6\pm0.9$ & $8.3\pm0.4$ & $1.2\pm0.5$ & \ding{55}\\
    J163103.4+395015 & $12\pm1$ & $2.7\pm0.6$ & $19\pm1$ & $7.8\pm0.4$ & $8.0\pm6.7$ & \ding{55}\\
    \multicolumn{1}{c}{}\\
    \hline
    \end{tabular}}
    \vspace{1mm}
    \caption*{(a) Name of the source; (b), (c), and (d) narrow ${\rm H}\alpha$, ${\rm H}\beta$, and $[{\rm O}III]5007\angstrom$ fluxes, respectively; (e) mass of the SMBH, estimated through the $M$--$\sigma^*$ relation; (f) bolometric luminosity of the AGN, estimated through \citet{duras2020} when the source had also an X-ray spectral analysis, and through \citet{netzer2009} otherwise; (g) check on the X-ray spectral analysis. Sources marked with \ding{51}/\ding{55} have/do not have an X-ray spectral analysis.}
\end{table*}

\section{Hardness ratio simulations}\label{sec:appendixb}

\begin{figure*}
    \centering
    \includegraphics[scale=0.6]{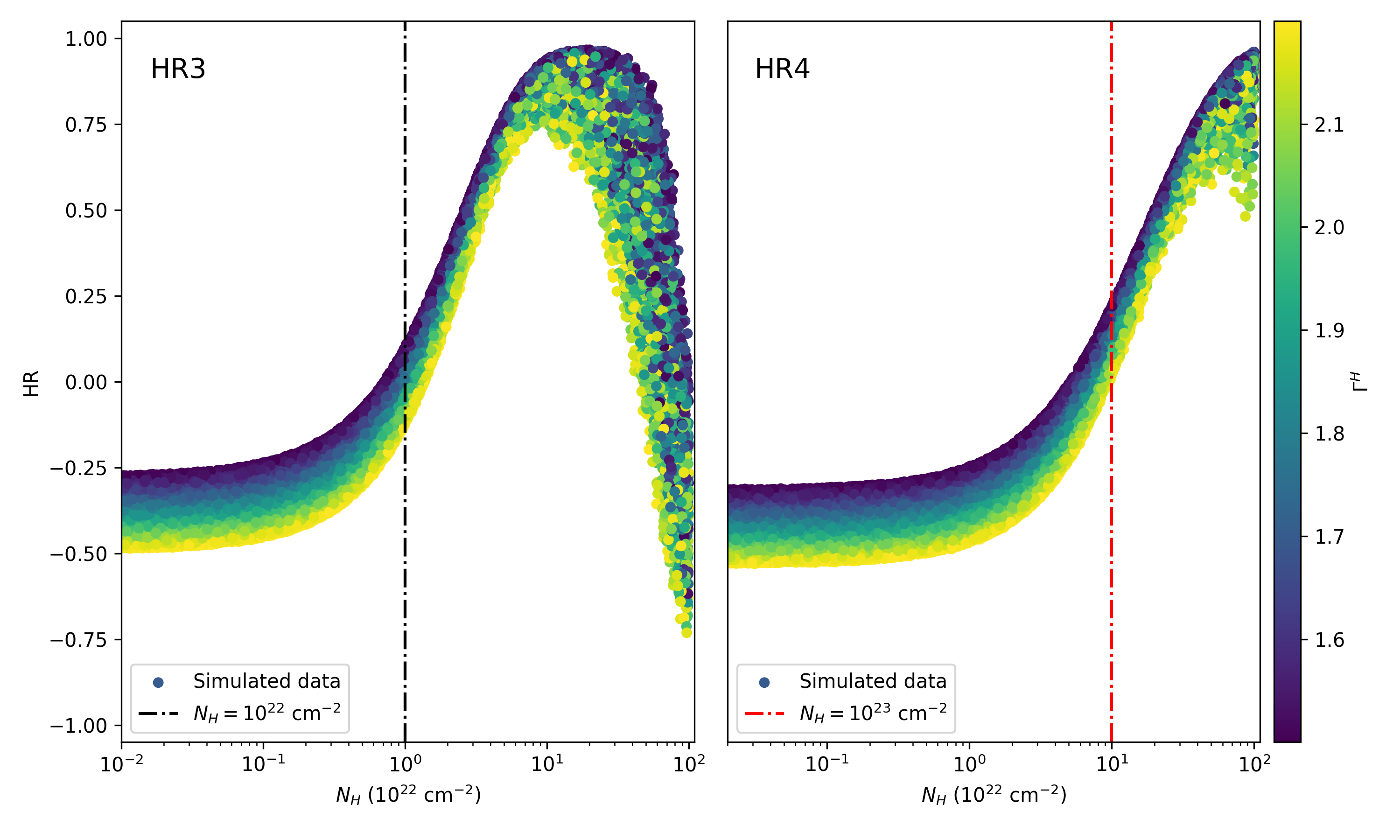}
    \caption{{\footnotesize Results from the HR3$-N_\mathrm{H}$ (left-hand panel) and HR4$-N_\mathrm{H}$ (right-hand panel) simulations. Data-points are colour-coded based on the $\Gamma^H$ value (colour bar on the right). The black and red, vertical, dashed lines represent the values $N_\mathrm{H}=10^{22}\unit{cm^{-2}}$ and $10^{23}\unit{cm^{-2}}$, respectively.}}
    \label{Figurea1}
\end{figure*}

The X-ray HR, defined as $(H-S)/(H+S)$, being $H$ and $S$ the counts per second in a hard and in a soft energy band, respectively, is strictly connected with $N_{\rm H}$ and its variability (e.g., \citealt{severgnini, derosa, cox2023}). Indeed, the higher the column density is, the more the primary emission of the AGN is suppressed, mainly decreasing the counts per second in the softer band, so that HR increases. Because of the different instrument characteristics, the bands in which HR is computed are $H=4.5$--$10\unit{keV}$, $S=2$--$4.5\unit{keV}$ (HR4), and $H=2$--$4.5\unit{keV}$, $S=1$--$2\unit{keV}$ (HR3) for XMM-Newton, and $H=2$--$8\unit{keV}$, $S=0.5$--$2\unit{keV}$ for Chandra. For XMM-Newton, HR4 is more sensitive to high obscuration ($N_{\rm H}>10^{23}\unit{cm^{-2}}$), while HR3 is more sensitive to moderate obscuration (e.g., the Compton-thin regime, $N_{\rm H}=10^{22}$--$10^{24}\unit{cm^{-2}}$). However, in any band, there is no unique correlation between HR and $N_{\rm H}$, since when $N_{\rm H}$ starts to enter the high absorbed regime, HR starts to decrease again (see Figure~\ref{Figurea1}).\\
\noindent We performed simulations with Xspec with the aim of deriving the correlation between XMM-Newton HR3 and HR4 with $N_\mathrm{H}$ (that we used to estimate the fraction of absorbed AGN in the PS). We set the number of simulations to 50,000, in order to have a high number of data-points. For each simulation, we generated a fake spectrum from which we extracted the counts per second in the bands of interest using the BLM described in Section~\ref{sec.3}. Each simulated spectrum is created using different values for $N_{\rm H}$, $\Gamma^H$, $\Gamma^S$, $K^H$, and $K^S$. $N_{\rm H}$ was varied in order to have data-points from $N_{\rm H}=10^{20}\unit{cm^{-2}}$ to $N_{\rm H}=10^{24}\unit{cm^{-2}}$. We decided to exclude $N_{\rm H}>10^{24}\unit{cm^{-2}}$, since in the CT regime the primary component of the AGN is totally suppressed in the observed X-ray bands (and, given the low redshift selection of the candidates, it would be not visible in XMM-Newton energy range). We varied the $\Gamma^H$ value in the range $1.5$--$2.2$, based on previous studies of AGN photon index distributions \citep{bianchi2, corral2011, ge2022}. In the hard and soft power law, $K^H$ and $K^S$ were varied in order to have $0.001<K^S/K^H<0.01$. We fixed the Milky Way hydrogen column density to $N^{\rm Gal}_{\rm H}=10^{20}\unit{cm^{-2}}$. Finally, the redshift was fixed to the mean value for the PS ($z\sim0.05$). This choice is justified with the fact that although HR evolves with redshift \citep{wang2004}, it does not depend on redshift in the range we are studying. The results of our simulations are shown in Figure~\ref{Figurea1}, where the degeneracy in the HR-$N_{\rm H}$ relation is clear, especially for HR3. Moreover, different values of $\Gamma^H$ and $K^S/K^H$ up to $N_{\rm H}=10^{24}\unit{cm^{-2}}$ produce different curves, as one can notice looking at the colour-code. Despite the previously described limitations when using HR as a proxy for $N_{\rm H}$, through our simulations we found that, depending on the $\Gamma^H$ value, a source enters the Compton-thin regime for HR3$\,\sim-0.2\ -\ 0.1$ (left-hand panel of Figure~\ref{Figurea1}) and the highly obscured regime ($N_\mathrm{H}>10^{23}\unit{cm^{-2}}$) for HR4$\,\sim0$--$0.25$ (right-hand panel of Figure~\ref{Figurea1}). Choosing a typical $\Gamma^H=1.9$, we can put a threshold on HR3$\,\geq0$ and HR4$\,\geq0.2$ to distinguish between un-obscured and obscured/highly obscured sources, respectively, and use this values of HR to estimate the fraction of Compton-thin isolated AGN in the PS.

 \section{Simulations for XMM-Newton and Chandra source detectability}\label{sec:appendixc}

\begin{figure}
    \centering
    \includegraphics[width=\linewidth]{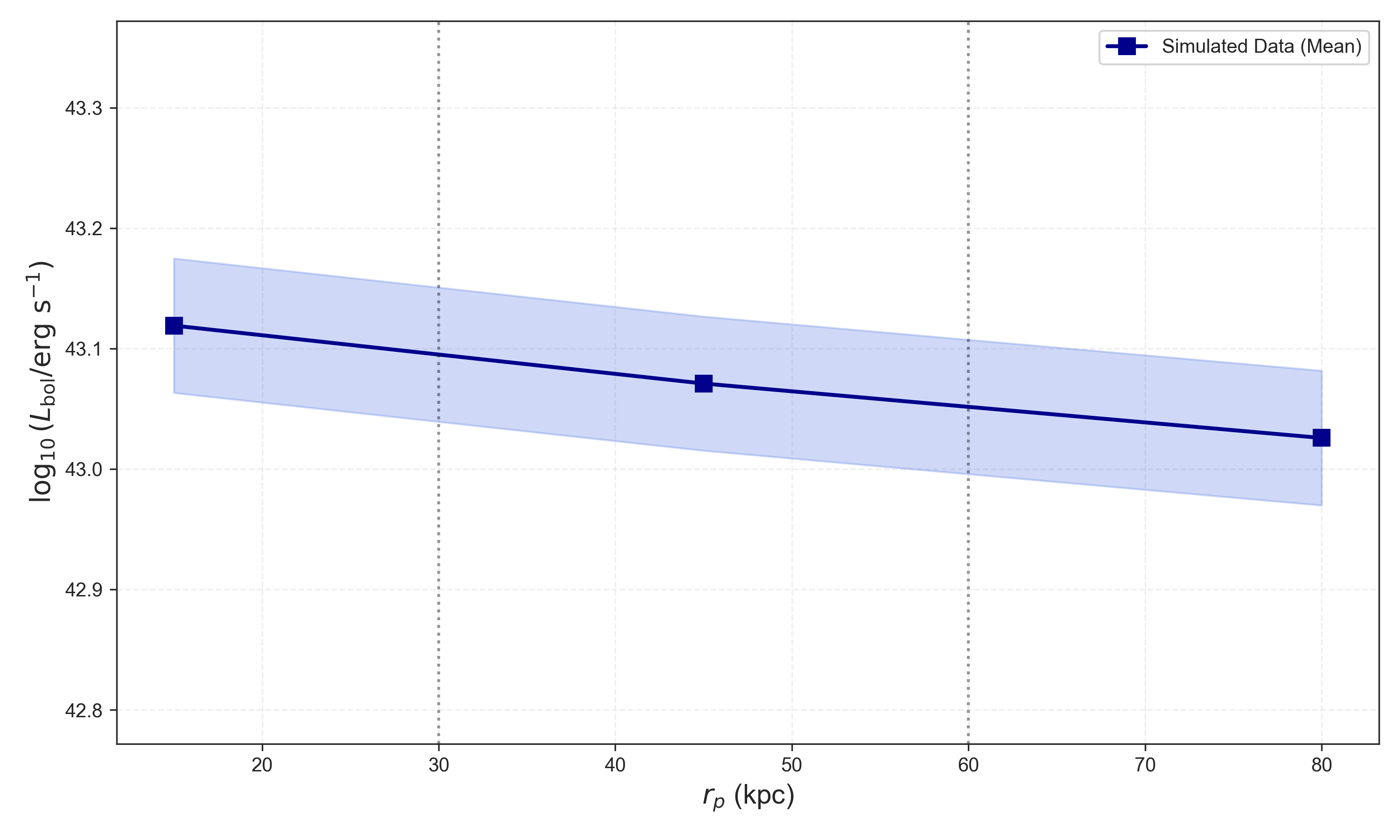}
    \caption{{\footnotesize Average simulated bolometric luminosity in each bin of projected spatial separation after the cuts on AGN fluxes. The shaded-blue region represents the errors on the mean bolometric luminosity in each bin.}}
    \label{figure_c1}
\end{figure}

{AGN in pairs at close separations (i.e., {$r_\mathrm{p}<30\unit{kpc}$}) were found to suffer from higher obscuration with respect to that in early merger stage (see Sections~\ref{sub.4.3}, \ref{sub.4.1}). This could mean that low-luminosity AGN may be not detected by XMM-Newton and {Chandra}, especially at the closest separations, if their X-ray fluxes fall below the detection limit of the available observations (see Section~\ref{sub.2.1}) due to a possible high obscuration. This could consequentially lead to a number of missing low-luminosity AGN at close separations, resulting in a potential bias for the correlation between {$\log_{10}(L_{\rm bol,2})$} and {$r_\mathrm{p}$}. In fact, if such kind of bias were confirmed, this would imply that the observed linear increase of {$\log_{10}(L_{\rm bol,2})$} with decreasing {$r_\mathrm{p}$} would be possibly driven by selection effects. To quantify how a selection bias would affect our dual-AGN sample, we performed simulations in {Xspec} to reproduce an unbiased population of AGN. To this goal, we simulated AGN spectra without imposing any constraint on the X-ray fluxes, but only requiring that each fake spectrum should satisfy the X-ray definition to be associated with an AGN source (i.e., intrinsic luminosity {$L(2-10\unit{keV})\geq10^{40}\unit{erg\ s^{-1}}$}). This ensured us to not exclude possible sources with X-ray fluxes below {$10^{-15}\unit{erg\ s^{-1}\ cm^{-2}}$} (for XMM-Newton) and {$10^{-16}\unit{erg\ s^{-1}\ cm^{-2}}$} (for {Chandra}), which are the possible missing AGN from our selection (see Section~\ref{sub.2.1}). For the simulations, we assumed as fitting model a simple absorbed power-law, {$f(E)=e^{-N_\mathrm{H}\sigma}KE^{-\Gamma^H}$} ({$ztbabs\times powerlaw$} in {Xspec}), varying the values of the photon index $\Gamma$ in the range {$1.5-2.2$}, according to previous studies on AGN photon index distributions \citep{bianchi2, corral2011, ge2022}. The normalisation {$K$} was instead varied in order to have {$10^{40}\unit{erg\ s^{-1}}\leq L(2-10\unit{keV})\leq10^{43}\unit{erg\ s^{-1}}$} (to match the intrinsic luminosities of our sample), depending on the simulated {$N_\mathrm{H}$} value (described below). We used this X-ray luminosity to estimate the bolometric luminosity of each AGN using \cite{duras2020}. The redshift values for the {$ztbabs$} component were varied in the range of our selection ({$0.01\lesssim z<0.1$}). We then included the effect of obscuration based on our results described in Section~\ref{sub.4.3}: we simulated three sets of data points in which we required the fraction of absorbed AGN to be {$70\%$}, {$70\%$}, and {$30\%$}, corresponding to the late, middle, and early stage of merger, respectively (see Figure~\ref{fig3}). The value of {$r_\mathrm{p}$} for each simulated AGN was randomly assigned in each bin of separation {$0-30\unit{kpc}$}, {$30-60\unit{kpc}$} and {$60-100\unit{kpc}$}, depending on the belonging set. The {$N_\mathrm{H}$} corresponding to absorbed AGN was varied in the range {$10^{22}\unit{cm^{-2}}\leq N_\mathrm{H}\leq10^{24}\unit{cm^{-2}}$}. The number of simulations was set to 500 for each bin of obscuration/separation both for XMM-Newton and {Chandra}, so that we had a total number of 3000 simulated data-points over all the separations, and an equal number of data-points in each bin. These simulations thus produced a complete population of AGN at all separations, that is unbiased against the absorption. Finally, we imposed flux cuts matching our selection criteria (see Section~\ref{sub.2.1}), in order to reproduce our observed AGN sample and include the possible obscuration bias. If a bias due to absorption exists, we thus expect that after the flux cut the number of low-luminosity AGN would be significantly reduced for the separations more affected by obscuration (i.e., the ones in the late state of merger), mimicking the observed trend between {$L_{\rm bol,2}$} and {$r_\mathrm{p}$} shown in the left-hand panel of Figure~\ref{fig5}. As a result, in Figure~\ref{figure_c1} is shown the average simulated bolometric luminosity in each bin of projected spatial separation, after the flux cut on the sources. A slight increase of the average bolometric luminosity with decreasing {$r_\mathrm{p}$} is visible (as expected after removing low-luminosity sources that are more distributed in the first and second bin), but this is only marginal (of the order of {$\sim0.1\unit{dex}$} between the early and late state of merger) and not statistically significant. As a further test, we also artificially increased the fraction of obscured AGN up to {$95\%$} in the bin {$0-30\unit{kpc}$} when performing the simulations, in order to consider a worst-case scenario of obscuration bias. Again, we did not find appreciable changing on the average bolometric luminosities between the early and late stage of merger, after the flux cuts. This means that the obscuration bias alone is not able to reproduce our observed trend between bolometric luminosity and separation of BH2 (see the upper, left-hand panel of Figure~\ref{fig6}). This is in agreement with the fact that a merger-driven physical mechanism is affecting the increase of bolometric luminosity for close pairs.}
 
\end{appendix}

\end{document}